\documentclass[seceq,preprint]{ptptex}

\newcommand{\sla}[1]{#1\hspace{-1.5mm}/}
\newcommand{\bsla}[1]{#1\hspace{-2.5mm}/}
\newcommand{\bbsla}[1]{#1\hspace{-3.0mm}/}
\newcommand{\dz}{\frac{dz}{2\pi i}}
\preprintnumber[3cm]{%<-- [..]: optional width of preprint # column.
YITP-06-67\\
KEK-TH-1123%\\PTPTeX ver.0.8
\\ December, 2006}
\title{%        %You can use \\ for explicit line-break
Lower-Dimensional Superstrings\\
in the Double-Spinor Formalism
}

\author{%       %Use \scshape  for the family name
Hiroshi \textsc{Kunitomo}$^1$
and
Shun'ya \textsc{Mizoguchi}$^{2,3}$%
}

\inst{%Affiliation, neglected when [addenda] or [errata]
$^1$Yukawa Institute for Theoretical Physics,
Kyoto University\\ Kyoto 606-8502, Japan\\%
$^2$High Energy Accelerator Research Organization (KEK)\\
Tsukuba 305-0801, Japan\\
$^3$Department of Particle and Nuclear Physics\\
The Graduate University for Advanced Studies\\
Tsukuba 305-0801, Japan
}

\abst{%         %this abstract is neglected when [addenda] or [errata]
We study lower-dimensional superstrings in the double-spinor 
formalism introduced by Aisaka and Kazama. 
These superstrings can be consistently quantized 
and are equivalent to the lower-dimensional 
pure-spinor superstrings proposed by Grassi and Wyllard.
The unexpected physical spectrum of the pure-spinor superstrings %can in fact 
may thus
be regarded as
a manifestation of  noncriticality.
We also discuss how to couple these covariant 
superstrings to the compactified degrees of freedom
described by the $N=2$ superconformal field theory.
}

\begin{document}

\maketitle

\section{Introduction}

The pure-spinor (PS) formalism, proposed by Berkovits, 
allows quantization of superstrings in a manner that preserves
the manifest super-Poincar\'e covariance.\cite{B}
Defined basically as a free conformal field theory (CFT), 
it provides a powerful framework for 
computing multipoint/multiloop amplitudes and for 
studying superstrings in RR backgrounds, which cannot be realized
with other formalisms.

One of the unsolved problems of the PS formalism is
to elucidate how the concept of ``criticality"
can be understood in this framework. 
In the RNS formalism, the critical dimension is that in which conformal anomalies 
cancel and the BRST charge becomes nilpotent, while in the light-cone Green-Schwarz (GS)
formalism, this is the only dimension in which the global Lorentz invariance is unbroken.
By contrast, 
the lower-dimensional versions of the PS formalism
have been constructed \cite{GW,W} by simply anticipating analogous free 
CFTs 
with some plausible ``pure spinor" conditions,\footnote{
It should be noted that 
this is in fact not the \textrm{pure-spinor} condition in lower dimensions.}
in which
the BRST charge is  exactly nilpotent and 
the Lorentz algebra has no anomaly.  Therefore, they seem to be 
consistent theories, even quantum mechanically, although they also have
some unexpected features such as the appearance of an \textit{off-shell} vector 
multiplet in the open string spectrum at the lowest level.
Like the original PS formalism, these theories are not based on
Lagrangians, and this makes it difficult to 
understand their fundamental nature, 
and whether or not (and if so how) they are 
related to lower-dimensional (non-critical) superstrings in other formalisms.

Recently, a Lagrangian formulation of the $(D=10)$
PS superstring was proposed by Aisaka and Kazama\cite{AK}. 
A special feature of their formulation is that it involves,
in addition to the ordinary superspace coordinates $(X^\mu,\theta^\alpha)$
in the GS superstring, another fermionic field,
$\tilde{\theta}^\alpha$. The Lagrangian of this double-spinor
(DS) formalism possesses world-sheet reparametrization invariance,
a manifest space-time super Poincar\'e symmetry
and a new local fermionic symmetry, which can be used to gauge away
one of the fermionic fields. 
Since the Lagrangian of the DS formalism is
reduced to the original GS Lagrangian  
by setting $\tilde{\theta}^\alpha$ to zero, this guarantees, 
at least classically, the equivalence of the DS superstring and
the GS superstring.

The DS superstring can be quantized without difficulty
using the conventional canonical BRST method. 
Surprisingly, the authors of \citen{AK} have shown that 
fields which have nontrivial Dirac brackets can be redefined in
such a way that all of them become free fields. Therefore, the quantized 
theory can be described by a simple conformal field theory represented 
by the free fields. It is also proved in \citen{AK} that the physical
spectrum of this DS superstring coincides with that of the PS superstring.

In this paper, to gain a better understanding of the mysterious features 
of the lower-dimensional PS superstrings, 
we study the corresponding lower-dimensional  (actually $d=4$ and $d=6$) 
DS superstrings. 
The equivalence to the lower-dimensional GS superstrings
is manifest for the same reason as in the ten-dimensional case.
We show that the lower-dimensional DS superstrings are also equivalent 
to the lower-dimensional PS superstrings.
Therefore, it is suggested that the unexpected physical spectrum
of the PS superstrings can be interpreted as
a manifestation of  ``noncriticality" in the PS formalism.
Then we search for a possible method for coupling these theories to 
the degrees of freedom of some compactified spaces,
which we call the Calabi-Yau (CY) sector. 
It is well known that such degrees of freedom can be described 
by unitary representations of the $N=2$ superconformal field 
theory with $c=9$ $(6)$ for $d=4$ $(6)$. 
We attempt to combine these two sectors in two different manners.
Unfortunately, however, the spectrum in the resultant 
combined system is a tensor product of the physical spectrum 
of two sectors and thus is not an expected on-shell spectrum. 

The paper is organized as follows. In \S\ref{PS} we briefly
review the $d=4$ and $d=6$ PS superstrings\cite{GW,W} 
defined by naive extensions of the ten-dimensional 
PS superstring. It is shown that the BRST charges 
are nilpotent without anomaly, but the lowest-level physical state
is an off-shell vector multiplet. 
The four-dimensional superstring in the DS formalism is 
studied in \S\ref{DS4}.
It is possible to carry out the quantization in a manner 
parallel to that for the ten-dimensional
case. The physical spectrum is shown to be equivalent
to the PS formalism in four dimensions. A similar argument
for the six-dimensional case is given in \S\ref{DS6}.
The coupling to the degrees of freedom of the compactified space
is discussed in \S\ref{CY}. We attempt to apply two different
methods for combining the two sectors, but the physical spectrum
is simply given by tensor products of the unexpected, off-shell,
spectrum of the lower-dimensional superstring with 
the (anti-)chiral ring of the CY sector.
Finally, in \S\ref{discuss}, we summarize our results and
discuss some remaining problems.
The appendix contains a summary of our notation and
spinor conventions.

\section{PS formalism in lower dimensions}\label{PS}

As in $D=10$,
the lower-dimensional PS superstring is defined by free fields
describing a map from the world-sheet to the target
superspace. The bosonic coordinates $X^\mu(z)$ 
satisfy the free field operator product\footnote{
In this section, 
we concentrate only on the holomorphic sector, or open strings.
It is easy to combine this sector with the anti-holomorphic 
sector to get closed strings.}
\begin{equation}
 X^\mu(z)X^\nu(w)\sim\eta^{\mu\nu}\log(z-w),
\end{equation}
where $\mu,\nu=0,1,\cdots,d-1$.
The fermionic coordinates are described by 
a first-order form of
conformal dimension $(1,0)$, whose explicit 
expression depends strongly on the space-time dimensionality.
The spinor conventions and the notation used in this paper
are summarized in the appendix \ref{conventions}.

\subsection{Four-dimensional PS superstring}

As is well known, the fermionic coordinates of
four-dimensional superspace consist of a pair of
complex conjugate Weyl spinors,
$\theta^\alpha$ and $\bar{\theta}^{\dot\alpha}$.
Introducing their conjugate fields,
$p_\alpha$ and $\bar{p}_{\dot\alpha}$, 
we assume the free field operator products
\begin{equation}
\theta^\alpha(z)p_\beta(w)\sim
\frac{\delta^\alpha_\beta}{z-w},\qquad
\bar{\theta}^{\dot\alpha}(z)\bar{p}_{\dot\beta}(w)
\sim\frac{\delta^{\dot\alpha}_{\dot\beta}}{z-w}. 
\end{equation}
We assign $\theta^\alpha$ and $\bar{\theta}^{\dot\alpha}$ 
conformal dimension 0
and $p_\alpha$ and $\bar{p}_{\dot\alpha}$ conformal dimension 1.
In addition to these fields, we also introduce 
the bosonic spinor 
fields $\lambda^\alpha$ and $\bar{\lambda}^{\dot\alpha}$ 
satisfying the \textit{pure-spinor constraint}
\begin{equation}
 \lambda\sigma^\mu\bar{\lambda}=0.\label{pure4}
\end{equation}
Using these pure-spinor fields, the BRST
charge is defined by
\begin{equation}
Q_{PS}=\oint\dz\left(\lambda^\alpha d_\alpha+
\bar{\lambda}^{\dot{\alpha}}\bar{d}_{\dot{\alpha}}\right).
\label{psbrs4}
\end{equation}
Here
$d_\alpha$ and $\bar{d}_{\dot{\alpha}}$ are 
the currents corresponding to the super-covariant
derivatives
\begin{subequations}\label{d4}
\begin{align}
d_\alpha=&p_\alpha-i\partial X^\mu(\sigma_\mu\bar{\theta})_\alpha
-\frac{1}{2}\left(
(\theta\sigma^\mu\partial\bar{\theta})
-(\partial\theta\sigma^\mu\bar{\theta})\right)
(\sigma_\mu\bar{\theta})_\alpha,\\
%%%%%%%%%%%
\bar{d}_{\dot\alpha}=&\bar{p}_{\dot\alpha}
-i\partial X^\mu(\theta\sigma_\mu)_{\dot\alpha}
-\frac{1}{2}\left(
(\theta\sigma^\mu\partial\bar{\theta})
-(\partial\theta\sigma^\mu\bar{\theta})\right)
(\theta\sigma_\mu)_{\dot\alpha},
\end{align}
\end{subequations}
which satisfy 
\begin{equation}
d_\alpha(z)\bar{d}_{\dot\alpha}(w)\sim
\frac{2i(\sigma_\mu)_{\alpha\dot\alpha}\pi^\mu(w)}{z-w},
\end{equation}
with
\begin{equation}
\pi^\mu=i\partial X^\mu+(\theta\sigma^\mu\partial\bar{\theta})
-(\partial\theta\sigma^\mu\bar{\theta}).\label{pi4}
\end{equation}

The BRST charge $Q_{PS}$ is nilpotent,
due to the pure-spinor constraint (\ref{pure4}),
even at the quantum level.
The physical states are defined as the cohomology of this
BRST charge. However, it possesses a rather unexpected spectrum,
because the on-shell condition is not imposed.
For example, the lowest-level vertex operator with ghost number 1,
which is expected to involve the physical state, has the form
\begin{equation}
 W=\lambda^\alpha A_\alpha(x,\theta,\bar{\theta})
+\bar{\lambda}^{\dot\alpha}A_{\dot{\alpha}}(x,\theta,\bar{\theta}),
\end{equation}
where $A_\alpha$ and $A_{\dot\alpha}$ are conventional superfields
of zero-modes $(x^\mu,\theta^\alpha,\bar{\theta}^{\dot\alpha})$.
Then the BRST invariance $\{Q,W\}=0$ implies  the conditions
\begin{subequations}
\begin{align}
& D_{(\alpha}A_{\beta)}(x,\theta,\bar{\theta})=0,\\
& \bar{D}_{(\dot\alpha}A_{\dot\beta)}(x,\theta,\bar{\theta})=0,\\
& D_\alpha A_{\dot\alpha}(x,\theta,\bar{\theta})
+\bar{D}_{\dot\alpha}A_\alpha(x,\theta,\bar{\theta})
+2i(\sigma^\mu)_{\alpha\dot\alpha}A_\mu(x,\theta,\bar{\theta})=0,
\end{align}
\end{subequations}
with an arbitrary superfield $A_\mu$.
These are the well-known torsion constraints\cite{WB}
and can be solved in terms of a real superfield
$V(x,\theta,\bar{\theta})$ as
\begin{equation}
 A_\alpha=iD_\alpha V,\qquad
 A_{\dot\alpha}=-i\bar{D}_{\dot\alpha}V,\qquad
 A_\mu=\frac{1}{4}(\bar\sigma_\mu)^{\dot{\alpha}\alpha}
[D_\alpha,\bar{D}_{\dot\alpha}]V,
\end{equation}
up to BRST exact pieces. There are no further restrictions,
and thus the lowest-level physical states constitute
an \textit{off-shell} vector multiplet.

\subsection{Six-dimensional PS superstring}

The fermionic coordinates of six-dimensional superspace 
are $SU(2)$-Majorana-Weyl (MW) spinors $\theta^\alpha_I$
$(I=1,2)$.
The free field operator product with the
conjugate field $p_\alpha^I$ is
\begin{equation}
\theta^\alpha_I(z)p_\beta^J(w)\sim
\frac{\delta^\alpha_\beta\delta_I^J}{z-w}.
\end{equation}
We assign $\theta^\alpha_I$ and $p_\alpha^I$
conformal dimension 0 and 1, respectively.
The bosonic pure-spinors $\lambda_\alpha^I$ satisfy
\begin{equation}
 \lambda_IC\gamma^\mu\lambda^I=0.
\end{equation}
Then the BRST charge is defined by
\begin{equation}
Q_{PS}=\oint\dz\lambda^\alpha_Id^I_\alpha,\label{psbrs6}
\end{equation}
which is also nilpotent quantum mechanically, since
the operator product relations
\begin{equation}
 d_\alpha^I(z)d_\beta^J(w)\sim
\frac{2i\epsilon^{IJ}(C\gamma_\mu)_{\alpha\beta}\pi^\mu(w)}
{z-w}
\end{equation}
hold between the super-covariant currents
\begin{subequations}\label{scc6}
\begin{align}
 d_\alpha^I=&p^I_\alpha+i\partial X^\mu(C\gamma_\mu\theta^I)_\alpha
+\frac{1}{2}(\theta^KC\gamma^\mu\partial\theta_K)
(C\gamma_\mu\theta^I)_\alpha,\label{d6}\\
\pi^\mu=&i\partial X^\mu+(\theta^KC\gamma^\mu\partial\theta_K).
\label{pi6}
\end{align}
\end{subequations}

The lowest-level vertex operator with ghost number 1
in six dimensions is now given by
\begin{equation}
 W=\lambda^\alpha_I A_\alpha^I(x,\theta_I),
\end{equation}
where $A_\alpha$ is a superfield of the zero-modes
$(x^\mu,\theta^\alpha_I)$. The BRST invariance $\{Q,W\}=0$
yields the conditions
\begin{subequations}
\begin{align}
& D_{(\alpha}^{(I}A^{J)}_{\beta)}=0,\\
& D_{I[\alpha}A_{\beta]}^I+
2i(C\gamma^\mu)_{\alpha\beta}A_\mu=0,
\end{align}
\end{subequations}
with an arbitrary superfield $A_\mu$.
These six-dimensional torsion constraints can also be solved
similarly to those in the four-dimensional case.\cite{HST,GW}
The lowest-level physical states are therefore also given by 
a six-dimensional \textit{off-shell} vector multiplet.

\section{DS formalism in four dimensions}\label{DS4}

The Lagrangian formulation of the PS superstring 
is given in \citen{AK}.
This formulation can be easily extended to the lower-dimensional case.

\subsection{Lagrangian, symmetries and constraints}

The four-dimensional superstring in the DS formalism
is defined by using the superspace coordinates 
$(x^\mu,\theta^{A\alpha},\bar{\theta}^{A\dot{\alpha}})$
and the additional fermionic fields
$(\tilde{\theta}^{A\alpha},\tilde{\bar{\theta}}^{A\dot{\alpha}})$
(A=1,2). The Lagrangian is then given by
\begin{subequations}\label{lag4}
\begin{align}
\mathcal{L}=&\mathcal{L}_K+\mathcal{L}_{WZ},\\
\mathcal{L}_K=&
-\frac{1}{2}\sqrt{-g}g^{ab}\Pi^\mu_a\Pi_{\mu b},\\
\mathcal{L}_{WZ}=&
\epsilon^{ab}\Pi^\mu_a(W_{\mu b}-\hat{W}_{\mu b})
-\epsilon^{ab}W^\mu_a\hat{W}_{\mu b},
\end{align}
\end{subequations}
where
\begin{align}
 \Pi^\mu_a=&\partial_aX^\mu
-\sum_{A=1}^2i\partial_a(\theta^A\sigma^\mu\tilde{\bar{\theta}}^A
-\tilde{\theta}^A\sigma^\mu\bar{\theta}^A)-\sum_{A=1}^2W^{A\mu}_a,\\
 W^{A\mu}_a=&i\Theta^A\sigma^\mu\partial_a\bar{\Theta}^A
-i\partial_a\Theta^A\sigma^\mu\bar{\Theta}^A,\\
\Theta^A=&\tilde{\theta}^A-\theta^A,\qquad
\bar{\Theta}^A=\tilde{\bar{\theta}}^A-\bar{\theta}^A.
\end{align}
We use the same notation as in Ref.~\citen{AK}; for example, 
$W^{1\mu}_a=W^\mu_a,\ W^{2\mu}_a=\hat{W}^\mu_a$, etc.
This Lagrangian is invariant under the world-sheet reparametrization
and the space-time Poincar\'e transformation, where the space-time
supersymmetry transformation is defined by 
\begin{subequations}\label{susy}
\begin{align}
\delta\theta^A=&\epsilon^A,\qquad
\delta\bar{\theta}^A=\bar{\epsilon}^A,\\
\delta\tilde{\theta}^A=&0,\qquad
\delta\tilde{\bar{\theta}}^A=0,\\
\delta X^\mu=&
\sum_{A=1}^2i(\epsilon^A\sigma^\mu\bar{\theta}^A
-\theta^A\sigma^\mu\bar{\epsilon}^A).
\end{align}
\end{subequations}
The additional fermionic fields are inert under
this supersymmetry.

The Lagrangian has another important symmetry,
the local supersymmetry, given by
\begin{subequations}
\begin{align}
\delta\theta^A=&\chi^A,\qquad \delta\tilde{\theta}^A=\chi^A,\\
\delta\bar{\theta}^A=&\bar{\chi}^A,\qquad 
\delta\tilde{\bar{\theta}}^A=\bar{\chi}^A,\\
\delta X^\mu=&
\sum_{A=1}^2i(\chi^A\sigma^\mu\bar{\Theta}^A
-\Theta^A\sigma^\mu\bar{\chi}^A), 
\end{align}
\end{subequations}
which guarantees the equivalence
to the conventional Green-Schwarz formalism.
Using this local symmetry, we can set the additional 
fermionic fields as $\tilde{\theta}^{A\alpha}=
\tilde{\bar{\theta}}^{A\dot{\alpha}}=0$. Then, the Lagrangian
(\ref{lag4}) becomes that of the Green-Schwarz 
formalism.\cite{GS}

Let us consider the canonical quantization of the Lagrangian
(\ref{lag4}). First, the canonical conjugate of $X^\mu$
can be computed as
\begin{equation}
 k_\mu=-\sqrt{-g}g^{0b}\Pi_{\mu b}+W_{\mu 1}-\hat{W}_{\mu 1}.
\end{equation}
Then, computing the canonical conjugates
$k^A_\alpha,\bar{k}^A_{\dot{\alpha}},\tilde{k}^A_\alpha$ and 
$\tilde{\bar{k}}^A_{\dot{\alpha}}$
of the fermionic fields $\theta^{A\alpha},\bar{\theta}^{A\dot{\alpha}},
\tilde{\theta}^{A\alpha}$ and
$\tilde{\bar{\theta}}^{A\dot{\alpha}}$,
we obtain the primary constraints
\begin{subequations}\label{primary4}
\begin{align}
D^A_\alpha=&k^A_\alpha+i(\bsla{k}\tilde{\bar{\theta}}^A)_\alpha
+i(k^\mu+\eta_A(\Pi^\mu_1+W^{\mu\bar{A}}_1))
(\sigma_\mu\bar{\Theta}^A)_\alpha\approx0,\\
%%%%%%%%%%%%%%%%%%%%%%%%%
\bar{D}^A_\alpha=&\bar{k}^A_{\dot{\alpha}}
+i(\tilde{\theta}^A\bsla{k})_{\dot{\alpha}}
+i(k^\mu+\eta_A(\Pi^\mu_1+W^{\mu\bar{A}}_1))
(\Theta^A\sigma_\mu)_{\dot{\alpha}}\approx0,\\
%%%%%%%%%%%%%%%%%%%%%%%%%
\tilde{D}^A_\alpha=&\tilde{k}^A_\alpha
-i(\bsla{k}\bar{\theta}^A)_\alpha
-i(k^\mu+\eta_A(\Pi^\mu_1+W^{\mu\bar{A}}_1))
(\sigma_\mu\bar{\Theta}^A)_\alpha\approx0,\\
%%%%%%%%%%%%%%%%%%%%%%%%%
\tilde{\bar{D}}^A_\alpha=&\tilde{\bar{k}}^A_{\dot{\alpha}}
-i(\theta^A\bsla{k})_{\dot{\alpha}}
-i(k^\mu+\eta_A(\Pi^\mu_1+W^{\mu\bar{A}}_1))
(\Theta^A\sigma_\mu)_{\dot{\alpha}}\approx0,
\end{align}
\end{subequations}
where $\eta_1=-\eta_2=1$ and $\bar{A}=2(1)$ for $A=1(2)$.
For later use, it is convenient to define
$\Delta_\alpha=D_\alpha+\tilde{D}_\alpha,\ 
\bar{\Delta}_{\dot{\alpha}}=\bar{D}_{\dot{\alpha}}
+\tilde{\bar{D}}_{\dot{\alpha}}$.

Using the ADM decomposition
\begin{equation}
g_{ab}=
\begin{pmatrix}
 -N^2+\gamma (N^1)^2 & \gamma N^1 \\
 \gamma N^1 & \gamma
\end{pmatrix} ,\label{ADM}
\end{equation}
we obtain the Hamiltonian as
\begin{align}
\mathcal{H}=&
\frac{N}{\sqrt{\gamma}}\frac{1}{2}\Big(
(k_\mu-W_{\mu1}+\hat{W}_{\mu1})(k^\mu-W^\mu_1+\hat{W}^\mu_1)
+\Pi^\mu_1\Pi_{\mu1}\Big)
\nonumber\\
&
+N^1(k_\mu-W_{\mu1}+\hat{W}_{\mu1})\Pi^\mu_1
+\dot{\theta}^{A\alpha}D^A_\alpha
+\dot{\bar{\theta}}^{A\dot{\alpha}}\bar{D}^A_{\dot{\alpha}}
+\dot{\tilde{\theta}}^{A\alpha}\tilde{D}^A_\alpha
+\dot{\tilde{\bar{\theta}}}^{A\dot{\alpha}}\tilde{\bar{D}}^A_{\dot{\alpha}},
\nonumber\\
\equiv&\frac{N}{\sqrt{\gamma}}T_0+N^1T_1
+\dot{\theta}^{A\alpha}D^A_\alpha
+\dot{\bar{\theta}}^{A\dot{\alpha}}\bar{D}^A_{\dot{\alpha}}
+\dot{\tilde{\theta}}^{A\alpha}\tilde{D}^A_\alpha
+\dot{\tilde{\bar{\theta}}}^{A\dot{\alpha}}\tilde{\bar{D}}^A_{\dot{\alpha}},
\label{hamiltonian4}
\end{align}
where the energy-momentum tensors are given by
\begin{subequations} 
\begin{align}
 T_+=&\frac{1}{2}(T_0+T_1)=\frac{1}{4}\Pi^\mu\Pi_\mu,\\
 T_-=&\frac{1}{2}(T_0-T_1)=\frac{1}{4}\hat{\Pi}^\mu\hat{\Pi}_\mu,
\end{align}
\end{subequations}
with
\begin{subequations} 
\begin{align}
 \Pi^\mu=&k^\mu-W^\mu_1+\hat{W}^\mu_1+\Pi^\mu_1,
\nonumber\\
=&k^\mu+X'^\mu-\sum_Ai(\theta^A\sigma^\mu\tilde{\bar{\theta}}^A
-\tilde{\theta}^A\sigma^\mu\bar{\theta}^A)'-2W^\mu_1,\\
%%%%%%%%%%%%%%%%%%%%%%%%%%%%
\hat{\Pi}^\mu=&k^\mu-W^\mu_1+\hat{W}^\mu_1-\Pi^\mu_1,
\nonumber\\
=&k^\mu-X'^\mu+\sum_Ai(\theta^A\sigma^\mu\tilde{\bar{\theta}}^A
-\tilde{\theta}^A\sigma^\mu\bar{\theta}^A)'+2\hat{W}^\mu_1. 
\end{align}
\end{subequations}

Next we set the canonical Poisson brackets as
\begin{subequations}\label{Poisson4}
\begin{align}
\{X^\mu(\sigma),k^\nu(\sigma')\}_P=&
\eta^{\mu\nu}\delta(\sigma-\sigma'),\\ 
%%%%%%%%%%%%%%%%%%%%%%
\{\theta^{A\alpha}(\sigma),k^B_\beta(\sigma')\}_P=&
-\delta^{AB}\delta^\alpha_\beta\delta(\sigma-\sigma'),\\ 
%%%%%%%%%%%%%%%%%%%%%%
\{\bar{\theta}^{A\dot{\alpha}}(\sigma),\bar{k}^B_{\dot{\beta}}(\sigma')\}_P=&
-\delta^{AB}\delta^{\dot{\alpha}}_{\dot{\beta}}\delta(\sigma-\sigma'),\\ 
%%%%%%%%%%%%%%%%%%%%%%
\{\tilde{\theta}^{A\alpha}(\sigma),\tilde{k}^B_\beta(\sigma')\}_P=&
-\delta^{AB}\delta^\alpha_\beta\delta(\sigma-\sigma'),\\ 
%%%%%%%%%%%%%%%%%%%%%%
\{\tilde{\bar{\theta}}^{A\dot{\alpha}}(\sigma),\tilde{\bar{k}}^B_{\dot{\beta}}(\sigma')\}_P=&
-\delta^{AB}\delta^{\dot{\alpha}}_{\dot{\beta}}\delta(\sigma-\sigma').
\end{align}
\end{subequations}
As in the ten-dimensional case,\cite{AK} the algebra of constraints
can be separated into left $(A=1$, or un-hatted) 
and right $(A=2$, or hatted) sectors, although the generators
include both un-hatted and hatted fields. 
Thus, for simplicity, we hereafter concentrate on the left sector.
Using the canonical Poisson brackets (\ref{Poisson4}), 
we can compute the Poisson brackets among
the fermionic constraint generators
$(\Delta_\alpha,\bar{\Delta}_{\dot{\alpha}},
\tilde{D}_\alpha,\tilde{\bar{D}}_{\dot{\alpha}})$ as
\begin{align}\label{const14}
\{\tilde{D}_\alpha(\sigma),\tilde{\bar{D}}_{\dot{\alpha}}(\sigma')\}_P=&
2i\Pi^\mu(\sigma)(\sigma_\mu)_{\alpha\dot{\alpha}}\delta(\sigma-\sigma'),
\end{align}
with all others vanishing.

In addition to these fermionic constraints, 
we also obtain the Virasoro constraint, $T_+\approx0$, 
for the left sector, from the Hamiltonian (\ref{hamiltonian4})
as the secondary constraint.
It is convenient to define the total energy-momentum tensor
\begin{align}
T=\frac{1}{4}\Pi^\mu\Pi_\mu+
\Theta'^\alpha\tilde{D}_\alpha
+\bar{\Theta}'^{\dot{\alpha}}\tilde{\bar{D}}_{\dot{\alpha}}
+\theta^\alpha\Delta_\alpha
+\bar{\theta}^{\dot{\alpha}}\bar{\Delta}_{\dot{\alpha}},
\end{align}
by adding the fermion contributions, which are also weakly vanishing,
due to the constraint (\ref{primary4}).
The Poisson bracket of this tensor is that of
the Virasoro algebra:
\begin{equation}
\{T(\sigma),T(\sigma')\}_P=
2T(\sigma)\delta'(\sigma-\sigma')
+T'(\sigma)\delta(\sigma-\sigma'). 
\end{equation}

Now, let us study the constraint algebra (\ref{const14}).
Due to the Virasoro constraints $T_+\approx0$,
this implies that half of the constraints 
$\tilde{D}_\alpha\approx0$ and 
$\tilde{\bar{D}}_{\dot{\alpha}}\approx0$ are first
class and the half are second class.
As is well known, these constraints cannot be separated covariantly,
and therefore we use the light-cone decomposition.
The Poisson bracket
\begin{equation}
\{\tilde{D}_1(\sigma),\tilde{\bar{D}}_{\dot1}(\sigma')\}_P
=2i\Pi^+(\sigma)\delta(\sigma-\sigma') \label{second4}
\end{equation}
shows that the first components $\tilde{D}_1$ and 
$\tilde{\bar{D}}_{\dot{1}}$ generate the second-class constraints.
The other half of the constraints come from the $\kappa$-symmetry 
generated by
\begin{subequations}
\begin{align}
K=&\tilde{D}_2-\frac{\Pi}{\Pi^+}\tilde{D}_1,\\
\bar{K}=&\tilde{\bar{D}}_{\dot2}-\frac{\bar{\Pi}}{\Pi^+}\tilde{\bar{D}}_{\dot1}, 
\end{align}
\end{subequations}
where $\Pi=\Pi^1+i\Pi^2$ and $\bar{\Pi}=\Pi^1-i\Pi^2$.
These are first class, because they satisfy the relations
\begin{equation}
\{K(\sigma),\bar{K}(\sigma')\}_P=
8i(\mathcal{T}+\mathcal{K})(\sigma)\delta'(\sigma-\sigma')+
(\tilde{D}_1,\tilde{\bar{D}}_{\dot{1}})\textrm{-terms}, 
\label{first4}
\end{equation}
where
\begin{subequations}
\begin{align}
\mathcal{T}=&\frac{T}{\Pi^+},\\
\mathcal{K}=&-\frac{1}{\Pi^+}(\Theta'^2K+\bar{\Theta}'^{\dot2}\bar{K}). 
\end{align}
\end{subequations}
These relations are closed up to 
$(\tilde{D}_1,\tilde{\bar{D}}_{\dot{1}})$-terms, which are
proportional to the second-class constraints
$\tilde{D}_1\approx0$ and $\tilde{\bar{D}}_{\dot1}\approx0$ 
and can be set to zero after calculating the Dirac bracket.
This is also consistent with the fact that the second-class 
constraints are transformed into themselves under this symmetry
generated by the first-class constraints:
\begin{subequations}
\begin{align}
\{K(\sigma),\tilde{D}_1(\sigma')\}_P=&
8i\frac{1}{\Pi^+}\bar{\Theta}'^{\dot2}\tilde{D}_1(\sigma)\delta(\sigma-\sigma'),\\
\{K(\sigma),\tilde{\bar{D}}_{\dot1}(\sigma')\}_P=&0,\\
\{\bar{K}(\sigma),\tilde{D}_1(\sigma')\}_P=&0,\\
\{\bar{K}(\sigma),\tilde{\bar{D}}_{\dot1}(\sigma')\}_P=&
8i\frac{1}{\Pi^+}\Theta'^2\tilde{\bar{D}}_{\dot1}(\sigma)\delta(\sigma-\sigma').
\end{align}
\end{subequations}

The above decomposition of the constraints explicitly breaks the manifest Lorentz
invariance. This breaking is, however, restricted to the sector of 
the additional fermionic fields, and therefore
the space-time supersymmetry defined by (\ref{susy}) can be linearly
realized. 

If we choose the semi-light-cone gauge, defined by
\begin{equation}
\tilde{\theta}^2\approx\tilde{\bar{\theta}}^{\dot{2}}\approx0,
\label{slcgauge}
\end{equation}
to fix the $\kappa$-symmetry,
all the constraints $\phi^I\approx0$ with
$\phi^I=(\tilde{D}_1,\tilde{\bar{D}}_{\dot{1}},
K,\bar{K},\tilde{\theta}^2,\tilde{\bar{\theta}}^{\dot{2}})$
become second class. We can take account of these
second-class constraints by computing the Dirac bracket
\begin{align}
 \{A(\sigma),B(\sigma')\}_D=&
\{A(\sigma),B(\sigma')\}_P
\nonumber\\
&
-\int d\sigma_1\{A(\sigma),\tilde{D}_1(\sigma_1)\}_P
\frac{1}{2i\Pi^+}(\sigma_1)
\{\tilde{\bar{D}}_{\dot1}(\sigma_1),B(\sigma')\}_P
\nonumber\\
&
-\int d\sigma_1\{A(\sigma),\tilde{\bar{D}}_{\dot1}(\sigma_1)\}_P
\frac{1}{2i\Pi^+}(\sigma_1)
\{\tilde{D}_1(\sigma_1),B(\sigma')\}_P 
\nonumber\\
&
+8i\int d\sigma_1\{A(\sigma),\tilde{\theta}^2(\sigma_1)\}_P
\mathcal{T}(\sigma_1)\{\tilde{\bar{\theta}}^{\dot2}(\sigma_1),B(\sigma')\}_P
\nonumber\\
&
+8i\int d\sigma_1\{A(\sigma),\tilde{\bar{\theta}}^{\dot2}(\sigma_1)\}_P
\mathcal{T}(\sigma_1)\{\tilde{\theta}^2(\sigma_1),B(\sigma')\}_P
\nonumber\\
&
+\int d\sigma_1\{A(\sigma),\tilde{\theta}^2(\sigma_1)\}_P
\{K(\sigma_1),B(\sigma')\}_P
\nonumber\\
&
+\int d\sigma_1\{A(\sigma),K(\sigma_1)\}_P
\{\tilde{\theta}^2(\sigma_1),B(\sigma')\}_P
\nonumber\\
&
+\int d\sigma_1\{A(\sigma),\tilde{\bar{\theta}}^{\dot2}(\sigma_1)\}_P
\{\bar{K}(\sigma_1),B(\sigma')\}_P
\nonumber\\
&
+\int d\sigma_1\{A(\sigma),\bar{K}(\sigma_1)\}_P
\{\tilde{\bar{\theta}}^{\dot2}(\sigma_1),B(\sigma')\}_P.
\end{align}

In this semi-light-cone gauge, the independent fields are
$(\tilde{\theta}^1,\tilde{\bar{\theta}}^{\dot1},X^\mu,k_\mu,
\theta^\alpha,k_\alpha,\bar{\theta}^{\dot{\alpha}},\bar{k}_{\dot\alpha})$,
which satisfy the Dirac brackets
\begin{subequations}
\begin{align}
\{\tilde{\theta^1}(\sigma),\tilde{\bar{\theta}}^{\dot1}(\sigma')\}_D=& 
\frac{i}{2\Pi^+}(\sigma)\delta(\sigma-\sigma'),\\
%%%%%%%%%%%%%%%%%%%%%%%%%%%%
\{X^\mu(\sigma),\tilde{\theta}^1(\sigma')\}_D=&
-\frac{1}{2\Pi^+}(\tilde{\theta}\sigma^\mu)_{\dot1}(\sigma)\delta(\sigma-\sigma'),\\
%%%%%%%%%%%%%%%%%%%%%%%%%%%%
\{X^\mu(\sigma),\tilde{\bar{\theta}}^{\dot1}(\sigma')\}_D=&
-\frac{1}{2\Pi^+}(\sigma^\mu\tilde{\bar{\theta}})_1(\sigma)\delta(\sigma-\sigma'),\\
%%%%%%%%%%%%%%%%%%%%%%%%%%%%
\{X^\mu(\sigma),X^\nu(\sigma')\}_D=&
\frac{i}{2\Pi^+}\left(
(\tilde{\theta}\sigma^\mu)_{\dot1}(\sigma^\nu\tilde{\bar{\theta}})_1
-(\tilde{\theta}\sigma^\nu)_{\dot1}(\sigma^\mu\tilde{\bar{\theta}})_1
\right)(\sigma)\delta(\sigma-\sigma'),\\
%%%%%%%%%%%%%%%%%%%%%%%%%%%%
\{X^\mu(\sigma),k^\nu(\sigma')\}_D=&
\eta^{\mu\nu}\delta(\sigma-\sigma')
\nonumber\\
&-\frac{i}{2\Pi^+}
\left((\tilde{\theta}\sigma^\mu)_{\dot1}(\sigma^\nu\bar{\Theta})_1
-(\Theta\sigma^\nu)_{\dot1}(\sigma^\mu\tilde{\bar{\theta}}_1)\right)(\sigma)
\delta'(\sigma-\sigma'),\\
%%%%%%%%%%%%%%%%%%%%%%%%%%%%
\{P^\mu(\sigma),k^\nu(\sigma')\}_D=&
-\frac{i}{2}\partial_\sigma\left(
\frac{1}{\Pi^+}\left(
(\Theta\sigma^\mu)_{\dot1}(\sigma^\nu\bar{\Theta})_1
-(\Theta\sigma^\nu)_{\dot1}(\sigma^\mu\bar{\Theta})_1
\right)(\sigma)\delta'(\sigma-\sigma')
\right),\\
%%%%%%%%%%%%%%%%%%%%%%%%%%%%
\{\tilde{\theta}^1(\sigma),k^\nu(\sigma')\}_D=&
-\frac{1}{2\Pi^+}(\Theta\sigma^\nu)_{\dot1}(\sigma)\delta'(\sigma-\sigma'),\\
%%%%%%%%%%%%%%%%%%%%%%%%%%%%
\{\tilde{\bar{\theta}}^{\dot1}(\sigma),k^\nu(\sigma')\}_D=&
-\frac{1}{2\Pi^+}(\sigma^\nu\bar{\Theta})_1(\sigma)\delta'(\sigma-\sigma').
\end{align}
\end{subequations}
These can be rewritten in simpler forms in
the constraint plane defined by 
$\tilde{\theta}^2=\tilde{\bar{\theta}}^{\dot2}=0$.
The non-trivial Dirac brackets are, for example,
\begin{subequations}
\begin{align}
\{\tilde{\theta^1}(\sigma),\tilde{\bar{\theta}}^{\dot1}(\sigma')\}_D=& 
\frac{i}{2\Pi^+}(\sigma)\delta(\sigma-\sigma'),\\
\{X^-(\sigma),\tilde{\theta}^1(\sigma')\}_D
=&-\frac{1}{\Pi^+}\tilde{\theta}^1 (\sigma)\delta(\sigma-\sigma'),\\
\{X^-(\sigma),\tilde{\bar{\theta}}^{\dot1}(\sigma')\}_D
=&-\frac{1}{\Pi^+}\tilde{\bar{\theta}}^{\dot1} (\sigma)\delta(\sigma-\sigma').
\end{align}
\end{subequations}

We have now separated, as in \citen{AK}, the first-class constraints from the 
second, the latter of which can be set to zero consistently with use of
the Dirac bracket. With their non-trivial forms, it may appear difficult 
to quantize the theory.
The authors of \citen{AK} found in $D=10$ that one can redefine 
the independent fields
so that the new ones satisfy simple free-field brackets. 
Here we show that we can do the same thing in the $D=4$ DS theory.

First, we define
\begin{equation}
S=\sqrt{2\Pi^+}\tilde{\theta}^1,\qquad
\bar{S}=\sqrt{2\Pi^+}\tilde{\bar{\theta}}^{\dot1}.
\end{equation}
Then $S$ and $\bar{S}$ become independent free fermions:
\begin{subequations}
\begin{align}
\{S(\sigma),\bar{S}(\sigma')\}_{D^*}=&i\delta(\sigma-\sigma'),\\ 
\{X^-(\sigma),S(\sigma')\}_{D^*}=&0,\\
\{X^-(\sigma),\bar{S}(\sigma')\}_{D^*}=&0.
\end{align}
\end{subequations}
Similarly for the remaining fields, 
we can find field redefinitions that make all of them
free fields:
\begin{subequations}
\begin{align}
P^\mu=&k^\mu-i(\tilde{\theta}\sigma^\mu\bar{\theta})'
+i(\theta\sigma^\mu\tilde{\bar{\theta}})'
+i(\hat{\tilde{\theta}}\sigma^\mu\hat{\bar{\theta}})'
-i(\hat{\theta}\sigma^\mu\hat{\tilde{\bar{\theta}}})',\\
%%%%%%%%%%%%%%%%%%%%%%%%%%%
p^A_1=&k^A_1+\eta^A
\left(
-iX'^+\tilde{\bar{\theta}}^{A\dot1}
-6(\theta^{A2}\bar{\theta}'^{A\dot2})\tilde{\bar{\theta}}^{A\dot1}
-2(\theta^{A2}\bar{\theta}^{A\dot2})\tilde{\bar{\theta}}'^{A\dot1}
\right) ,\\
%%%%%%%%%%%%%%%%%%%%%%%%%%%
p^A_2=&k^A_2+\eta^A
\Big(
-iX'\tilde{\bar{\theta}}^{A\dot1}
-6(\bar{\theta}'^{A\dot2}\theta^{A1})\tilde{\bar{\theta}}^{A\dot1}
+2(\bar{\theta}^{A\dot2}\tilde{\theta}^{A1})'\tilde{\bar{\theta}}^{A\dot1}
\nonumber\\
&
+3(\bar{\theta}^A\bar{\theta}^A)'\tilde{\theta}^{A1}
+2\bar{\theta}^{A\dot2}(\tilde{\bar{\theta}}^{A\dot1}\tilde{\theta}^{A1})'
-2(\bar{\theta}^{A\dot2}\theta^{A1})\tilde{\bar{\theta}}'^{A\dot1}
+2(\bar{\theta}^A\bar{\theta}^A)\tilde{\theta}'^{A1}
\Big) ,\\
%%%%%%%%%%%%%%%%%%%%%%%%%%%%
\bar{p}^A_{\dot1}=&\bar{k}^A_{\dot1}+\eta^A
\left(
-iX'^+\tilde{\theta}^{A\dot1}
-6(\bar{\theta}^{A\dot2}\theta'^{A2})\tilde{\theta}^{A1}
-2(\bar{\theta}^{A\dot2}\theta^{A2})\tilde{\theta}'^{A1}
\right) ,\\
%%%%%%%%%%%%%%%%%%%%%%%%%%%%
\bar{p}^A_{\dot2}=&\bar{k}^A_{\dot2}+\eta^A
\Big(
-i\bar{X}'\tilde{\theta}^{A1}
-6(\theta'^{A2}\bar{\theta}^{A\dot1})\tilde{\theta}^{A1}
+2(\theta^{A2}\tilde{\bar{\theta}}^{A\dot1})'\tilde{\theta}^{A1}
\nonumber\\
&
+3(\theta^A\theta^A)'\tilde{\bar{\theta}}^{A\dot1}
+2\theta^{A2}(\tilde{\theta}^{A1}\tilde{\bar{\theta}}^{A\dot1})'
-2(\theta^{A2}\bar{\theta}^{A\dot1})\tilde{\theta}'^{A1}
+2(\theta^A\theta^A)\tilde{\bar{\theta}}'^{A\dot1}
\Big),
\end{align}
\end{subequations}
which satisfy
\begin{subequations}
\begin{align}
\{X^\mu(\sigma),P^\nu(\sigma')\}_D=&\eta^{\mu\nu}\delta(\sigma-\sigma'),\\
\{\theta^{A\alpha}(\sigma),p^B_\beta(\sigma')\}_D
=&-\delta^{AB}\delta^\alpha_\beta\delta(\sigma-\sigma'),\\
\{\bar{\theta}^{A\dot\alpha}(\sigma),\bar{p}^B_{\dot\beta}(\sigma')\}_D
=&-\delta^{AB}\delta^\alpha_\beta\delta(\sigma-\sigma').
\end{align}
\end{subequations}
with the remaining brackets vanishing.
These redefinitions also yield a complete separation\cite{AK}
of the left and right sectors for the fundamental quantities,
including the constraint generators. For instance, we have
\begin{subequations}
\begin{align}
\Pi^\mu=&P^\mu+{X^\mu}'
\nonumber\\
&-2i(\theta\sigma^\mu\bar{\theta}') 
+2i(\theta'\sigma^\mu\bar{\theta}) 
-2i(\tilde{\theta}\sigma^\mu\tilde{\bar{\theta}}') 
+2i(\tilde{\theta}'\sigma^\mu\tilde{\bar{\theta}})
+4i(\tilde{\theta}\sigma^\mu\bar{\theta}') 
-4i(\theta'\sigma^\mu\tilde{\bar{\theta}}),\\
%%%%%%%%%%%%%%%%%%%%%%%%%%%%%%%%%%%
\hat{\Pi}^\mu=&P^\mu-{X^\mu}'
\nonumber\\
&+2i(\hat{\theta}\sigma^\mu\hat{\bar{\theta}}') 
-2i(\hat{\theta}'\sigma^\mu\hat{\bar{\theta}}) 
+2i(\hat{\tilde{\theta}}\sigma^\mu\hat{\tilde{\bar{\theta}}}') 
-2i(\hat{\tilde{\theta}}'\sigma^\mu\hat{\tilde{\bar{\theta}}})
-4i(\hat{\tilde{\theta}}\sigma^\mu\hat{\bar{\theta}}') 
+4i(\hat{\theta}'\sigma^\mu\hat{\tilde{\bar{\theta}}}). 
\end{align}
\end{subequations}

Now let us rewrite the constraint generators in terms of 
these free fields. They become
\begin{subequations} 
\begin{align}
D_1=&d_1+i\sqrt{2\Pi^+}\bar{S},\\
D_2=&d_2+i\sqrt{\frac{2}{\Pi^+}}\Pi\bar{S}
+\frac{4}{\Pi^+}S\bar{S}(\bar{\theta}^{\dot2})',\\
\bar{D}_{\dot1}=&\bar{d}_{\dot1}+i\sqrt{2\Pi^+}S,\\
\bar{D}_{\dot2}=&\bar{d}_{\dot2}+i\sqrt{\frac{2}{\Pi^+}}\bar{\Pi}S
-\frac{4}{\Pi^+}S\bar{S}(\theta^2)',\\
\mathcal{T}=&-\frac{1}{4}\frac{\Pi^\mu\Pi_\mu}{\Pi^+},
\end{align}
\end{subequations} 
where $d_\alpha$ and $\bar{d}_{\dot{\alpha}}$ are the 
supercovariant spinor conjugates
\begin{subequations} 
\begin{align}
 d_\alpha&=p_\alpha-i(P^\mu+{X^\mu}')(\sigma_\mu\bar{\theta})_\alpha
+\left[(\theta'\sigma^\mu\bar{\theta})-(\theta\sigma^\mu\bar{\theta}')\right]
(\sigma_\mu\bar{\theta})_\alpha,\\
%%%%%%%%%%%%%%%%%%%%%%
 \bar{d}_{\dot{\alpha}}&=\bar{p}_{\dot{\alpha}}
-i(P^\mu+{X^\mu}')(\theta\sigma_\mu)_{\dot{\alpha}}
+\left[(\theta'\sigma^\mu\bar{\theta})-(\theta\sigma^\mu\bar{\theta}')\right]
(\theta\sigma_\mu)_{\dot{\alpha}}.
\end{align}
\end{subequations}

\subsection{Quantization}

Since we have redefined all the fields as free fields,
there is no difficulty in the quantization.
We only need to replace the Dirac bracket by the quantum bracket as
\begin{subequations}
\begin{align}
[X^\mu(\sigma),P^\nu(\sigma')]=&i\eta^{\mu\nu}\delta(\sigma-\sigma'),\\
\{p^A_\alpha(\sigma),\theta^{B\beta}(\sigma')\}=&
-i\delta^{AB}\delta_\alpha^\beta\delta(\sigma-\sigma'),\\
\{\bar{p}^A_{\dot{\alpha}}(\sigma),\bar{\theta}^{B\dot{\beta}}(\sigma')\}=&
-i\delta^{AB}\delta_{\dot{\alpha}}^{\dot{\beta}}\delta(\sigma-\sigma'),\\
\{S(\sigma),\bar{S}(\sigma')\}=&-\delta(\sigma-\sigma').
\end{align}
\end{subequations}
These can be translated into the OPE relations 
in the Euclidean formulation with radial quantization.
In order to obtain the standard normalization, we set
$T=1/2\pi\alpha'=1$ and make the replacement
$P^\mu+{X^\mu}'\rightarrow 2i\partial X^\mu$,
$p_\alpha\rightarrow 2ip_\alpha$, $\bar{p}_{\dot{\alpha}}\rightarrow
2i\bar{p}_{\dot{\alpha}}$, $S\rightarrow\sqrt{2}iS$ and
$\bar{S}\rightarrow-\sqrt{2}i\bar{S}$, where 
$\partial$ represents $\partial/\partial z$. The OPEs for the basic fields
of the left (holomorphic) sector then become
\begin{subequations}
\begin{align}
 X^\mu(z)X^\nu(w)\sim&\eta^{\mu\nu}\log(z-w),\\
p_\alpha(z)\theta^\beta(w)\sim&\frac{\delta_\alpha^\beta}{z-w},\\
\bar{p}_{\dot\alpha}(z)\bar{\theta}^{\dot\beta}(w)
\sim&\frac{\delta_{\dot\alpha}^{\dot\beta}}{z-w},\\
S(z)\bar{S}(w)\sim&\frac{1}{z-w}.
\end{align}
\end{subequations}
It is also convenient to change the normalizations as
\begin{equation}
 D_\alpha\rightarrow \frac{1}{2i}D_\alpha,\qquad
\bar{D}_{\dot{\alpha}}\rightarrow \frac{1}{2i}\bar{D}_{\dot{\alpha}},\qquad
\Pi\rightarrow \frac{1}{2}\Pi,
\end{equation}
and to rescale $\mathcal{T}$ so that $\mathcal{T}=-(1/2)\Pi^\mu\Pi_\mu/\Pi^+$.

Using these rescaled free fields, 
we rewrite the constraints in such a manner that separates the
non-covariant fermions $S$ and $\bar{S}$ from the remaining covariant
sector. Using the super-covariant current $(\ref{pi4})$, $\Pi^\mu$ can be
written as
\begin{subequations}
\begin{align}
\Pi^+=&\pi^+,\\
\Pi^-=&\pi^-+\frac{1}{\pi^+}\left(S\partial\bar{S}-\partial S\bar{S}\right)
-2i\sqrt{\frac{2}{\pi^+}}\left(
S\partial\bar{\theta}^{\dot1}+\partial\theta^1\bar{S}\right),\\
\Pi=&\pi+2i\sqrt{\frac{2}{\pi^+}}S\partial\bar{\theta}^{\dot2},\\
\bar{\Pi}=&\bar{\pi}+2i\sqrt{\frac{2}{\pi^+}}\partial\theta^2\bar{S}.
\end{align}
\end{subequations}
The constraint generators are then rewritten 
in terms of the free fields as
\begin{subequations}
\begin{align}
D_1=&d_1-i\sqrt{2\pi^+}\bar{S},\\
%%%%%%%%%%%%%%%%%%%%%%%%%
D_2=&d_2-i\sqrt{\frac{2}{\pi^+}}\pi\bar{S}
-\frac{2}{\pi^+}S\bar{S}\partial\bar{\theta}^{\dot 2},\\
%%%%%%%%%%%%%%%%%%%%%%%%%
\bar{D}_{\dot 1}=&\bar{d}_{\dot 1}+i\sqrt{2\pi^+}S,\\ 
%%%%%%%%%%%%%%%%%%%%%%%%%
\bar{D}_{\dot 2}=&\bar{d}_{\dot 2}+i\sqrt{\frac{2}{\pi^+}}\bar{\pi}S
+\frac{2}{\pi^+}S\bar{S}\partial\theta^2,\\
%%%%%%%%%%%%%%%%%%%%%%%%%
\mathcal{T}=&
-\frac{1}{2}\frac{\pi^\mu\pi_\mu}{\pi^+}
-\frac{1}{2}\frac{S\partial\bar{S}}{\pi^+}
+\frac{1}{2}\frac{\partial S\bar{S}}{\pi^+}
+i\sqrt{\frac{2}{\pi^+}}(S\partial\bar{\theta}^{\dot 1}
+\partial\theta^1\bar{S})
\nonumber\\
&
+i\sqrt{\frac{2}{(\pi^+)^3}}
\left(\bar{\pi}S\partial\bar{\theta}^{\dot2}
+\pi\partial\theta^2\bar{S}\right)
+4\frac{S\bar{S}\partial\theta^2\partial\bar{\theta}^{\dot2}}{(\pi^+)^2},
\end{align}
\end{subequations}
where $d_\alpha$ and $\bar{d}_{\dot\alpha}$ are super-covariant currents (\ref{d4}).

The constraint algebras forms by the generators above are not closed in
their present forms. However, these expressions are classical, although they
have normal ordering ambiguities in general. 
What is particularly noteworthy in the $D=10$ DS theory is that 
these ambiguities are used to modify the generators in such a way that 
their algebras become closed. A similar happens here:
Including such ``quantum corrections," the constraint generators become
\begin{subequations}\label{qc4}
\begin{align}
\widehat{D}_1=&D_1,\qquad
\widehat{\bar{D}}_{\dot1}=\bar{D}_{\dot1},\\
\widehat{D}_2=&D_2-\frac{\partial^2\bar{\theta}^{\dot2}}{\pi^+}
+\frac{1}{2}\frac{\partial\pi^+\partial\bar{\theta}^{\dot2}}{(\pi^+)^2},\\
\widehat{\bar{D}}_{\dot2}=&\bar{D}_{\dot2}
-\frac{\partial^2\theta^2}{\pi^+}
+\frac{1}{2}\frac{\partial\pi^+\partial\theta^2}{(\pi^+)^2},\\
\widehat{\mathcal{T}}=&\mathcal{T}+
\frac{\partial\theta^2\partial^2\bar{\theta}^{\dot2}}{(\pi^+)^2}
-\frac{\partial^2\theta^2\partial\bar{\theta}^{\dot2}}{(\pi^+)^2}
-\frac{1}{8}\frac{\partial^2\log\pi^+}{\pi^+},
\end{align}
\end{subequations}
and then they form a closed first-class  algebra: 
\begin{align}
\widehat{D}_2(z)\widehat{\bar{D}}_{\dot2}(w)\sim&
\frac{4\widehat{\mathcal{T}}(w)}{z-w},
\end{align}
with all others vanishing.
Using these generators of the first-class constraints,
one can straightforwardly construct the BRST charge
according to the conventional procedure as
\begin{equation}
\tilde{Q}=\oint\frac{dz}{2\pi i}\left(
\tilde{\lambda}^\alpha\widehat{D}_\alpha
+\tilde{\bar{\lambda}}^{\dot\alpha}\widehat{\bar{D}}_{\dot\alpha}
+c\widehat{\mathcal{T}}
-4\tilde{\lambda}^2\tilde{\bar{\lambda}}^{\dot2}b
\right),\label{BRSilde4}
\end{equation}
where $\tilde{\lambda}^\alpha$ and 
$\tilde{\bar{\lambda}}^{\dot{\alpha}}$
are \textit{unconstrained} bosonic spinor ghosts. 
Fermionic ghosts $b$ and $c$ satisfying
$b(z)c(w)\sim1/(z-w)$ are also introduced.
This is exactly nilpotent quantum mechanically,
due to the correction (\ref{qc4}).

As in the ten-dimensional case, we can construct the composite
$B$-ghost
\begin{equation}
 B=b\pi^+-\tilde{\omega}_\alpha\partial\theta^\alpha
-\tilde{\bar{\omega}}_{\dot{\alpha}}\partial\bar{\theta}^{\dot{\alpha}},
\label{b4}
\end{equation}
in terms of which the energy-momentum tensor can be expressed as
\begin{align}
\{\tilde{Q},B(z)\}=&T(z)\nonumber\\
=&-\frac{1}{2}\pi^\mu\pi_\mu-\frac{1}{8}\partial^2\log\pi^+
-\frac{1}{2}S\partial\bar{S}+\frac{1}{2}\partial S\bar{S}
\nonumber\\
&-d_\alpha\partial\theta^\alpha
-\bar{d}_{\dot{\alpha}}\partial\bar{\theta}^{\dot{\alpha}}
-\tilde{\omega}_\alpha\partial\tilde{\lambda}^\alpha
-\tilde{\bar{\omega}}_{\dot{\alpha}}\partial\tilde{\bar{\lambda}}
^{\dot{\alpha}}
-b\partial c,
\end{align}
where the bosonic anti-ghosts 
$\tilde{\omega}_\alpha$ and $\tilde{\bar{\omega}}_{\dot{\alpha}}$
are introduced as conjugates of $\tilde{\lambda}^\alpha$
and $\tilde{\bar{\lambda}}^{\dot{\alpha}}$ satisfying
\begin{equation}
\tilde{\lambda}^\alpha(z)\tilde{\omega}_\beta(w)
\sim\frac{\delta^\alpha_\beta}{z-w},
\qquad
\tilde{\bar{\lambda}}^{\dot{\alpha}}(z)\tilde{\bar{\omega}}_{\dot{\beta}}(w)
\sim\frac{\delta^{\dot{\alpha}}_{\dot{\beta}}}{z-w}. 
\end{equation} 
This energy-momentum tensor has vanishing 
central charge, due to the second term involving $\log\pi^+$.
We can rescale the fermionic $bc$-ghosts to obtain the conventional
reparametrization ghosts with conformal dimensions $2$ and $-1$ as
\begin{align}
e^RTe^{-R}
=&-\frac{1}{2}\pi^\mu\pi_\mu+\frac{7}{8}\partial^2\log\pi^+
-\frac{1}{2}S\partial\bar{S}+\frac{1}{2}\partial S\bar{S}
\nonumber\\
&-d_\alpha\partial\theta^\alpha
-\bar{d}_{\dot{\alpha}}\partial\bar{\theta}^{\dot{\alpha}}
-\tilde{\omega}_\alpha\partial\tilde{\lambda}^\alpha
-\tilde{\bar{\omega}}_{\dot{\alpha}}\partial\tilde{\bar{\lambda}}
^{\dot{\alpha}}
-\partial bc-2b\partial c
\end{align}
by carrying out
the similarity transformation generated by
\begin{equation}
R=-\oint\dz bc\log\pi^+.
\end{equation}

\subsection{Equivalence to the PS formalism in four dimensions}

Finally, we show that the cohomology of the BRST charge $\tilde{Q}$ 
(\ref{BRSilde4}) in the DS formalism is the same
as that of the PS formalism (\ref{psbrs4}).
To begin with, we have to explicitly solve 
the four-dimensional pure-spinor constraint, which
can also be written as
\begin{equation}
\lambda^\alpha\bar{\lambda}^{\dot{\alpha}}=0. \label{psc4}
\end{equation}
This has the solutions $\lambda^\alpha=0$ and 
$\bar{\lambda}^{\dot\alpha}=0$.\footnote{
If $\bar{\lambda}^{\dot\alpha}$ is the complex conjugate of
$\lambda^\alpha$, the pure-spinor constraint (\ref{psc4})
has no non-trivial solution. This is a common difficulty
in the PS formalism, and to avoid it we consider them to be
independent fields in this section.}

As the first step to prove the equivalence, we show 
the decoupling of the fermionic ghost pair $(b,c)$ and 
one of the two bosonic ghost pairs
$(\tilde{\lambda}^2,\tilde{\omega}_2)$ and 
$(\tilde{\bar{\lambda}}^{\dot2},\tilde{\bar{\omega}}_{\dot2})$.
We have to consider the two cases $\tilde{\lambda}^2\ne0$ and
$\tilde{\lambda}^{\dot2}\ne0$ separately. These correspond to 
the two branches of solutions of the pure-spinor constraint. 
The total Hilbert space is the union of  these 
two cases.

Let us first consider the case $\tilde{\lambda}^2\ne0$
corresponding to the branch $\bar{\lambda}^{\dot\alpha}$. 
In this case, we can consider the similarity
transformation generated by
\begin{equation}
X=-\frac{1}{4}\oint\frac{dz}{2\pi i}\frac{c\widehat{\bar{D}}_{\dot2}}
{\tilde{\lambda}^2},
\end{equation}
which transforms the BRST charge $\tilde{Q}$ into
\begin{equation}
e^X\tilde{Q}e^{-X}=\delta_b+Q^{(1)},\label{brs4}
\end{equation}
where
\begin{align}
\delta_b=&-4\oint\frac{dz}{2\pi i}\tilde{\lambda}^2\tilde{\bar{\lambda}}
^{\dot2}b,\\
Q^{(1)}=&\oint\frac{dz}{2\pi i}
(\tilde{\lambda}^\alpha
 \widehat{D}_\alpha+\tilde{\bar{\lambda}}^{\dot1}\widehat{\bar{D}}_{\dot1}).
\end{align}
The general argument of homological perturbation theory\cite{AKhp}
shows that the cohomology of $\tilde{Q}$ coincides with that of
$Q^{(1)}$ in the Hilbert space without 
$b$, $c$, $\tilde{\bar{\lambda}}^{\dot2}$
and $\tilde{\bar{\omega}}_{\dot2}$ decoupled as a quartet, 
due to $\delta_b$.\footnote{
This is due to the fact that
the cohomology of $\delta_b$ must satisfy a part of the
pure-spinor constraint (\ref{psc4})
$\tilde{\lambda}^2\tilde{\bar{\lambda}}^{\dot2}=0$.
This yields $\tilde{\bar{\lambda}}^{\dot2}=0$, that is,
the decoupling of $\tilde{\bar{\lambda}}^{\dot2}$,
since we consider the case $\tilde{\lambda}^2\ne0$.} 

We further carry out the second transformation, which
consists of two successive similarity transformations.
The first transformation is given by
$e^YQ^{(1)}e^{-Y}$, with
\begin{equation}
Y=-\frac{1}{2}\oint\frac{dz}{2\pi i}S\bar{S}\log\pi^+. 
\end{equation}
This yields the replacement
\begin{equation}
S\rightarrow\frac{S}{\sqrt{\pi^+}},\qquad 
\bar{S}\rightarrow\sqrt{\pi^+}\bar{S},
\end{equation}
which also implies  a shift  of  the conformal weight of 
$(S,\bar{S})$  from $(1/2,1/2)$ to $(1,0)$.
This transformation results in the decoupling of
$S$, $\bar{S}$, $\tilde{\bar{\lambda}}^{\dot1}$ and 
$\tilde{\bar{\omega}}_{\dot1}$ as a quartet. 
Indeed, the BRST charge is transformed as
\begin{align}
&Q^{(2)}=e^YQ^{(1)}e^{-Y}=\delta+Q+d,\\
&\delta=\sqrt{2}i\oint\frac{dz}{2\pi i}\bar{\lambda}^{\dot1}S,\qquad
Q=\oint\frac{dz}{2\pi i}\lambda^\alpha d_\alpha,\\
d=&\oint\frac{dz}{2\pi i}\left(
-\lambda^2\frac{\partial^2\bar{\theta}^{\dot2}}{\pi^+}
+\lambda^2\frac{\partial\pi^+\partial\bar{\theta}^{\dot2}}{(\pi^+)^2}
+\bar{\lambda}^{\dot1}\bar{d}_{\dot1}-\sqrt{2}i\lambda^1\pi^+\bar{S}
-\sqrt{2}i\lambda^2\pi \bar{S}
\right),
\end{align}
and the form of $\delta$ indicates such a decoupling.
Subsequently, the transformation
\begin{align}
&e^ZQ^{(2)}e^{-Z}=\delta+Q,\label{brsf4}\\
&\delta=\sqrt{2}i\oint\frac{dz}{2\pi i}\bar{\lambda}^{\dot1}S,\\
&Q=\oint\frac{dz}{2\pi i}\lambda^\alpha d_\alpha,\label{brs4f1}
\end{align}
with
\begin{equation}
 Z=\oint\frac{dz}{2\pi i}\left(
\frac{i}{\sqrt{2}}\bar{d}_{\dot1}\bar{S}
+\frac{\partial\theta^2\partial\bar{\theta}^{\dot2}}{\pi^+}\right),
\end{equation}
completes the second transformation. Now the final BRST charge
(\ref{brsf4}) has the same form as the first transformation (\ref{brs4}).
The cohomology of $Q^{(1)}$ coincides with that of $Q$ 
in the Hilbert space without 
$S$, $\bar{S}$, $\tilde{\bar{\lambda}}^{\dot1}$
and $\tilde{\bar{\omega}}_{\dot1}$ decoupled as a quartet,
due to $\delta$. This BRST charge $Q$ is identical to that of the
four-dimensional PS superstring in the branch $\bar{\lambda}^{\dot\alpha}=0$.

We can similarly consider the other branch,
$\tilde{\bar{\lambda}}^{\dot2}\ne0$.
The first similarity transformation in this case is generated by
\begin{equation}
 \bar{X}=-\frac{1}{4}\oint\frac{dz}{2\pi i}\frac{c\widehat{D}_2}
{\bar{\lambda}^{\dot2}},
\end{equation}
which transforms the BRST charge as
\begin{align}
e^{\bar{X}}\tilde{Q}e^{-\bar{X}}=&
\delta_b+Q^{(1)},\\
\delta_b=&-4\oint\frac{dz}{2\pi i}\lambda^2\bar{\lambda}^{\dot2}b,\\
Q^{(1)}=&\oint\frac{dz}{2\pi i}
(\lambda^1\widehat{D}_1+\bar{\lambda}^{\dot\alpha}\widehat{\bar{D}}
_{\dot\alpha}).
\end{align}
Then $\lambda^2,\ \omega_2,\ c$ and $b$ are decoupled as a quartet.

The second similarity transformation, given by
\begin{subequations}
\begin{align}
\bar{Y}=&\frac{1}{2}\oint\frac{dz}{2\pi i}S\bar{S}\log\pi^+, \\
 \bar{Z}=&-\oint\frac{dz}{2\pi i}\left(
\frac{i}{\sqrt{2}}d_1S
+\frac{\partial\theta^2\partial\bar{\theta}^{\dot2}}{\pi^+}\right),
\end{align}
\end{subequations}
leads to
\begin{align}
&e^{\bar{Z}}e^{\bar{Y}}Q^{(1)}e^{-\bar{Y}}e^{-\bar{Z}}
=\delta+Q,\\
&\delta=-\sqrt{2}i\oint\frac{dz}{2\pi i}\lambda^1\bar{S},\\
&Q=\oint\frac{dz}{2\pi i}\bar{\lambda}^{\dot\alpha}\bar{d}_{\dot\alpha},
\label{brs4f2}
\end{align}
and $\lambda^1,\ \omega_1,\ S$ and $\bar{S}$ are 
decoupled as a quartet. The BRST charge $Q$ is that of the
four-dimensional PS superstring in the branch $\lambda^\alpha=0$.

Thus the physical states are finally obtained as
the union of the cohomologies of the BRST charges (\ref{brs4f1})
and (\ref{brs4f2}). They coincide with those of the
four-dimensional PS superstring defined by the cohomology
of the BRST charge with the constrained spinor (\ref{psbrs4}).

\section{DS formalism in six dimensions}\label{DS6}

Similarly to the previous section, in this section
we consider the six-dimensional superstring in the DS formalism.

\subsection{Lagrangian, symmetries and constraints}

The six-dimensional superstring in the DS formalism
is defined using superspace coordinates $(x^\mu,\theta^{A\alpha}_I)$
and the additional fermionic field $\tilde{\theta}^{A\alpha}_I$ $(A=1,2)$.
Using these fundamental fields, the Lagrangian is given by
$\mathcal{L}=\mathcal{L}_K+\mathcal{L}_{WZ}$, with
\begin{subequations} 
\begin{align}
\mathcal{L}_K=&
-\frac{1}{2}\sqrt{-g}g^{ab}\Pi^\mu_a\Pi_{\mu b},\\
\mathcal{L}_{WZ}=&
\epsilon^{ab}\Pi^\mu_a(W_{\mu b}-\hat{W}_{\mu b})
-\epsilon^{ab}W^\mu_a\hat{W}_{\mu b}, 
\end{align}
\end{subequations}
where
\begin{align}
 \Pi^\mu_a=&\partial_aX^\mu
-\sum_{A=1}^2i\partial_a(\theta^{IA}C\gamma^\mu\tilde{\theta}^A_I)
-\sum_{A=1}^2W^{A\mu}_a,\\
 W^{A\mu}_a=&i(\Theta^{IA}C\gamma^\mu\partial_a\Theta^A_I),\\
\Theta^A_I=&\tilde{\theta}^A_I-\theta^A_I.
\end{align}
This Lagrangian is invariant under the world-sheet reparametrization and
the space-time Poincar\'e transformation, where the space-time
supersymmetry is defined by
\begin{subequations}
\begin{align}
\delta\theta^A_I=&\epsilon^A_I,\qquad
\delta\tilde{\theta}^A_I=0,\\
\delta X^\mu=&
\sum_{A=1}^2i(\epsilon^{IA}C\gamma^\mu\theta^A_I).
\end{align}
\end{subequations}
In six dimensions, the local supersymmetry is given by
\begin{subequations}
\begin{align}
\delta\theta^A_I=&\chi^A_I,\qquad \delta\tilde{\theta}^A_I=\chi^A_I,\\
\delta X^\mu=&
\sum_{A=1}^2i(\chi^{IA}C\gamma^\mu\Theta^A_I),
\end{align}
\end{subequations}
which guarantees the equivalence to the conventional Green-Schwarz
formalism. 

The canonical conjugate of $X^\mu$ is obtained as
\begin{equation}
 k_\mu=-\sqrt{-g}g^{0b}\Pi_{\mu b}+W_{\mu 1}-\hat{W}_{\mu 1}.
\end{equation}
By computing the canonical conjugates 
$k^{IA}_\alpha$ and $\tilde{k}^{IA}_\alpha$
of the fermionic fields $\theta^{\alpha A}_I$ and
$\tilde{\theta}^{\alpha A}_I$, we obtain the primary constraints
\begin{subequations}
\begin{align}
D^{IA}_\alpha=&p^{IA}_\alpha
-iP^\mu(C\gamma_\mu\tilde{\theta}^{IA})_\alpha
-i(P^\mu+\eta^A(\Pi^\mu_1+W^{\mu A'}_1))(C\gamma_\mu\Theta^{IA})_\alpha
\approx0,\\
%%%%%%%%%%%%%%%%%%%%%%%%%
\tilde{D}^{IA}_\alpha=&\tilde{p}^{IA}_\alpha
+iP^\mu(C\gamma_\mu\theta^{IA})_\alpha
+i(P^\mu+\eta^A(\Pi^\mu_1+W^{\mu A'}_1))(C\gamma_\mu\Theta^{IA})_\alpha
\approx0,
\end{align}
\end{subequations}
where $\eta_1=-\eta_2=1$ and $ A'=2(1)$ for $A=1(2)$.
We also define the generators
$\Delta^{IA}_\alpha=D^{IA}_\alpha+\tilde{D}^{IA}_\alpha$.

Using the same ADM decomposition of the world-sheet metric
as in the four-dimensional case (\ref{ADM}),
we can obtain the Hamiltonian as
\begin{align}
\mathcal{H}=&
\frac{N}{\sqrt{\gamma}}\frac{1}{2}\Big(
(P_\mu-W_{\mu1}+\hat{W}_{\mu1})(P^\mu-W^\mu_1+\hat{W}^\mu_1)
+\Pi^\mu_1\Pi_{\mu1}\Big)
\nonumber\\
&
+N^1(P_\mu-W_{\mu1}+\hat{W}_{\mu1})\Pi^\mu_1
+\sum_A\dot{\theta}^A_ID^{IA}
+\sum_A\dot{\tilde{\theta}}^A_I\tilde{D}^{IA},
\nonumber\\
=&\frac{N}{\sqrt{\gamma}}T_0+N^1T_1
+\sum_A\dot{\theta}^A_ID^{IA}
+\sum_A\dot{\tilde{\theta}}^A_I\tilde{D}^{IA}.
\end{align}
Here, the energy-momentum tensors are given by
\begin{subequations}
\begin{align}
 T_+=&\frac{1}{2}(T_0+T_1)=\frac{1}{4}\Pi^\mu\Pi_\mu,\\
 T_-=&\frac{1}{2}(T_0-T_1)=\frac{1}{4}\hat{\Pi}^\mu\hat{\Pi}_\mu,
\end{align}
\end{subequations}
with
\begin{subequations}
\begin{align}
\Pi^\mu=&k^\mu-W^\mu_1+\hat{W}^\mu_1+\Pi^\mu_1,\nonumber\\
=&k^\mu+X'^\mu
-\sum_Ai(\theta^{IA}C\gamma^\mu\tilde{\theta}^A_I)'
-2W^\mu_1,\\
\hat{\Pi}^\mu=&k^\mu-W^\mu_1+\hat{W}^\mu_1-\Pi^\mu_1,\nonumber\\
=&k^\mu-X'^\mu
+\sum_Ai(\theta^{IA}C\gamma^\mu\tilde{\theta}^A_I)'
+2\hat{W}^\mu_1. 
\end{align}
\end{subequations}

Then we can set the canonical Poisson brackets as
\begin{subequations}
\begin{align}
\{X^\mu(\sigma),P^\nu(\sigma')\}_P=&
\eta^{\mu\nu}\delta(\sigma-\sigma'),\\ 
\{\theta^{A\alpha}_I(\sigma),p^{JB}_\beta(\sigma')\}_P=&
-\delta^{AB}\delta^J_I\delta^\alpha_\beta\delta(\sigma-\sigma'),\\ 
\{\tilde{\theta}^{A\alpha}_I(\sigma),\tilde{p}^{JB}_\beta(\sigma')\}_P=&
-\delta^{AB}\delta_I^J\delta^\alpha_\beta\delta(\sigma-\sigma'). 
\end{align}
\end{subequations}
Using these canonical Poisson brackets, we can calculate
the constraint algebra of the fermionic generators
$(\Delta^I_\alpha,\tilde{D}^I_\alpha)$ as
\begin{align}
 \{\tilde{D}^I_\alpha(\sigma),\tilde{D}^J_\beta(\sigma')\}_P
=&-2i\epsilon^{IJ}(C\gamma_\mu)_{\alpha\beta}\Pi^\mu(\sigma)
\delta(\sigma-\sigma'),\label{const6}
\end{align}
with all others vanishing.

The total energy-momentum tensor is now given by
\begin{equation}
T=T_++t_+
=\frac{1}{4}\Pi^\mu\Pi_\mu+{\Theta'}_I^\alpha\tilde{D}^I_\alpha,
\end{equation}
which forms the Virasoro algebra:
\begin{equation}
 \{T_+(\sigma),T_+(\sigma')\}_P=
2T_+(\sigma)\delta'(\sigma-\sigma')
+T'(\sigma)\delta(\sigma-\sigma').
\end{equation}

We can decompose the constraint $\tilde{D}^I_\alpha\approx0$ 
into the second-class constraint $\tilde{D}^I_a\approx0$
and the first-class constraint $K^I_{\dot{a}}\approx0$ as
\begin{equation}
K^I_{\dot{a}}=\tilde{D}^I_{\dot{a}}
+\frac{(\epsilon\bbsla{\Pi}\ )_{\dot{a}b}}{\Pi^+}
\tilde{D}^{Ib}.
\end{equation}
These $\tilde{D}^I_a$ satisfy
\begin{subequations}
\begin{align}
&\{\tilde{D}^I_a(\sigma),\tilde{D}^J_b(\sigma')\}_P
=-2i\epsilon^{IJ}\epsilon_{ab}\Pi^+(\sigma)\delta(\sigma-\sigma'), \\
%%%%%%%%%%%%%%%%%%%%%%%%
&\{K^I_{\dot{a}}(\sigma),K^J_{\dot{b}}(\sigma')\}_P=
-8i\epsilon^{IJ}\epsilon_{\dot{a}\dot{b}}
(\mathcal{T}+\mathcal{K})(\sigma)\delta'(\sigma-\sigma')+
\tilde{D}\textrm{-terms},\label{KK}
\end{align}
\end{subequations}
where
$(\epsilon\bbsla{\Pi}\ )_{\dot{a}b}=(\epsilon\gamma_i)_{\dot{a}b}\Pi^i$,
$\mathcal{T}=\frac{T}{\Pi^+}$ and 
$\mathcal{K}=\frac{K^K_{\dot{c}}\Theta'^{\dot{c}}_K}{\Pi^+}$.
The $\tilde{D}$-terms in (\ref{KK}) can be set equal
to zero after taking the Dirac bracket.

By choosing the semi-light-cone gauge 
$\tilde{\theta}^{\dot{a}}_I\approx0$ 
for the first-class constraint, 
we finally have three second-class constraints:
\begin{equation}
\tilde{D}^I_a\approx0,\qquad
K^I_{\dot{a}}\approx0,\qquad
\tilde{\theta}^{\dot{a}}_i\approx0.
\end{equation}
The Dirac bracket for these second-class constraints
is defined by
\begin{align}
 \{A(\sigma),B(\sigma')\}_{D}=&
\{A(\sigma),B(\sigma')\}_P
\nonumber\\
&
+\int d\sigma_1\{A(\sigma),\tilde{D}^I_a(\sigma_1)\}_P
\frac{\epsilon_{IJ}\epsilon^{ab}}{2i\Pi^+}(\sigma_1)
\{\tilde{D}^J_b(\sigma_1),B(\sigma')\}_P
\nonumber\\
&
+\int d\sigma_1\{A(\sigma),K^I_{\dot{a}}(\sigma_1)\}_P
\{\tilde{\theta}^{\dot{a}}_I(\sigma_1),B(\sigma')\}_P
\nonumber\\
&
+\int d\sigma_1\{A(\sigma),\tilde{\theta}^{\dot{a}}_I(\sigma_1)\}_P
\{K^I_{\dot{a}}(\sigma_1),B(\sigma')\}_P
\nonumber\\
&
-8i\int d\sigma_1 \{A(\sigma),\tilde{\theta}^{\dot{a}}_I(\sigma_1)\}_P
\epsilon^{IJ}\epsilon_{\dot{a}\dot{b}}\mathcal{T}(\sigma_1)
\{\tilde{\theta}^{\dot{b}}_J(\sigma_1),B(\sigma')\}_P.
\end{align}
In this semi-light cone gauge,
the independent fields are
$(X^\mu,P_\mu,\theta^\alpha_I,p^I_\alpha,\tilde{\theta}^a_I)$,
which on the constraint plane 
$\tilde{\theta}^{\dot{a}}_I\approx0$ satisfy, for example,
\begin{subequations}
\begin{align}
&\{\tilde{\theta}^a_I(\sigma),\tilde{\theta}^b_J(\sigma')\}_D=
-\frac{i\epsilon_{IJ}\epsilon^{ab}}{2\Pi^+}(\sigma)\delta(\sigma-\sigma'),\\
%%%%%%%%%%%%
&\{X^-(\sigma),\tilde{\theta}^a_I(\sigma')\}_D=
-\frac{1}{\Pi^+}\tilde{\theta}^a_I(\sigma)\delta(\sigma-\sigma'),\\
%%%%%%%%%%%%
&\{X^-(\sigma),P^-(\sigma')\}_D=
\frac{2i}{\Pi^+}\tilde{\theta}^a_I\theta^I_a(\sigma)\delta'(\sigma-\sigma'),\\
%%%%%%%%%%%%
&\{X^-(\sigma),P^i(\sigma')\}_D=
-\frac{i}{\Pi^+}\tilde{\theta}^a_I(\epsilon\bar{\gamma}^i)_{a\dot{b}}
\theta^{I\dot{b}}(\sigma)\delta'(\sigma-\sigma'),\\
%%%%%%%%%%%%
&\{\tilde{\theta}^a_I(\sigma),P^-(\sigma')\}_D=
-\frac{1}{\Pi^+}\Theta^a_I(\sigma)\delta'(\sigma-\sigma'),\\
%%%%%%%%%%%%
&\{\tilde{\theta}^a_I(\sigma),P^i(\sigma')\}_D=
\frac{1}{2\Pi^+}{(\bar{\gamma}^i)^a}_{\dot{c}}\theta^{\dot{c}}_I(\sigma)
\delta(\sigma-\sigma').
\end{align}
\end{subequations} 
We can again redefine the fields so that the new ones become free.
We first redefine the non-covariant fermionic field as
\begin{equation}
S^a_I=\sqrt{2\Pi^+}\tilde{\theta}^a_I.
\end{equation}
Then $S^a_I$ satisfies the free relations
\begin{subequations}
\begin{align}
\{S^a_I(\sigma),S^b_J(\sigma')\}_D=&
-i\epsilon_{IJ}\epsilon^{ab}\delta(\sigma-\sigma'),\\ 
\{X^-(\sigma),S^a_I(\sigma')\}_D=&0.
\end{align}
\end{subequations}
Similarly, we can obtain the free fields through the redefinitions
\begin{subequations}
\begin{align}
P^\mu=&k^\mu-i(\tilde{\theta}^KC\gamma^\mu\theta_K)'
+i(\hat{\tilde{\theta}}^KC\gamma^\mu\hat{\theta}_K)',\\
%%%%%%%%%%%%%%%%%%%%%%%%%%%
p^{IA}_a=&k^{IA}_a+\eta^A
\left(
-iX'^+\tilde{\theta}^{IA}_a
-4\theta^{KA}_{\dot{c}}\theta'^{\dot{c}A}_K\tilde{\theta}^{IA}_a
+2\theta^{IA}_{\dot{c}}(\theta^{\dot{c}A}_K\tilde{\theta}^{KA}_a)'
+2(\theta^{IA}_{\dot{c}}\theta^{\dot{c}A}_K)'\tilde{\theta}^{KA}_a
\right),\\
%%%%%%%%%%%%%%%%%%%%%%%%%%%
p^{IA}_{\dot{a}}=&k^{IA}_{\dot{a}}+\eta^A
\Big(
i(\epsilon\bbsla{X}\ ')_{\dot{a}c}\tilde{\theta}^{IcA}
-2\theta^{IA}_{\dot{a}}\theta^{KA}_c\tilde{\theta}'^{cA}_K
-2\theta^{KA}_{\dot{a}}\theta^{IA}_c\tilde{\theta}'^{cA}_K
-6\theta'^{IA}_{\dot{a}}\theta^{KA}_c\tilde{\theta}^{cA}_K
\nonumber\\
&\hspace{20mm}
-6\theta^{KA}_{\dot{a}}\theta'^{IA}_c\tilde{\theta}^{cA}_K
-2\theta^{IA}_{\dot{a}}\tilde{\theta}^{KA}_c\tilde{\theta}'^{cA}_K
+4\theta^{KA}_{\dot{a}}\tilde{\theta}^{IA}_c\tilde{\theta}'^{cA}_K
-2(\theta^{K A'}_{\dot{a}}\tilde{\theta}^{ A'}_{Kc})'\tilde{\theta}^{IcA}
\Big).
\end{align}
\end{subequations}
These redefined fields satisfy
\begin{subequations}
\begin{align}
\{X^\mu(\sigma),P^\nu(\sigma')\}_D=&\eta^{\mu\nu}
\delta(\sigma-\sigma'),\\ 
\{\theta^{A\alpha}_I(\sigma),p^{BJ}_\beta(\sigma')\}_D=&
-\delta^{AB}\delta_I^J\delta^\alpha_\beta\delta(\sigma-\sigma'),
\end{align}
\end{subequations}
with the remaining brackets vanishing.

\subsection{Quantization}

The quantization is obtained by replacing the Dirac bracket
with the (anti-)commutation relations, which can be rewritten
in the form of the operator product expansion in the conventional
radial quantization, after the appropriate field rescalings:
\begin{subequations}
\begin{align}
 X^\mu(z)X^\nu(w)\sim&\eta^{\mu\nu}\log(z-w),\\
p^I_\alpha(z)\theta^\beta_J(w)\sim&
\frac{\delta^I_J\delta_\alpha^\beta}{z-w},\\
S^a_I(z)S^b_J(w)\sim&-\frac{\epsilon_{IJ}\epsilon^{ab}}{z-w}.
\end{align}
\end{subequations}

The constraint generators are classically given by
\begin{subequations}
\begin{align}
D^I_a=&d^I_a+\sqrt{2\pi^+}S^I_a,\\
%%%%%%%%%%%%%%%%%%%%%%%%%
D^I_{\dot{a}}=&d^I_{\dot{a}}
-\sqrt{\frac{2}{\pi^+}}(\epsilon\sla{\pi})_{\dot{a}b}S^{Ib}
+\frac{2}{\pi^+}S^I_cS^c_K\partial\theta^K_{\dot{a}},
\nonumber\\
=&
d^I_{\dot{a}}
-\sqrt{\frac{2}{\pi^+}}(\epsilon\sla{\pi})_{\dot{a}b}S^{Ib}
-\frac{1}{\pi^+}
(\epsilon\gamma_i)_{\dot{a}b}S^{Ib}
S^{Kc}(\epsilon\bar{\gamma}_i)_{c\dot{d}}\partial\theta^{\dot{d}}_K,\\
%%%%%%%%%%%%%%%%%%%%%%%%%
\mathcal{T}=&
-\frac{1}{2}\frac{\pi^\mu\pi_\mu}{\pi^+}
-\frac{1}{2}\frac{S^K_a\partial S^a_K}{\pi^+}
-\sqrt{\frac{2}{\pi^+}}\partial\theta^K_aS^a_K
\nonumber\\
&\hspace{20mm}
-\sqrt{\frac{2}{\pi^+}}
\frac{\partial\theta^{K\dot{a}}(\epsilon\bsla{\pi})_{\dot{a}b}S^b_K}{\pi^+}
-2\frac{\partial\theta^{K\dot{a}}
\partial\theta^L_{\dot{a}}S^b_KS_{Lb}}{(\pi^+)^2},
\end{align}
\end{subequations}
where the super-covariant currents $d^I_\alpha$ and $\pi^\mu$
are defined by (\ref{scc6}).
Including ``quantum" corrections, as we have done in $D=4$, they become%
\footnote{
We should note that, in contrast to the $D=4$ case, the extra terms in 
Eqs. (\ref{6dDadot}) and (\ref{6dT}) \textit{cannot} be regarded as 
coming solely from the normal ordering ambiguities; the origin 
of these terms is unclear. In any case, however,  the addition of these
terms closes the algebra and leads to a consistent theory, as we show below.
}
\begin{subequations}
\begin{align}
\widehat{D}^I_a=&D^I_a,\\
\widehat{D}^I_{\dot{a}}=&D^I_{\dot{a}}
-2\frac{\partial^2\theta^I_{\dot{a}}}{\pi^+}
+\frac{\partial\pi^+\partial\theta^I_{\dot{a}}}{(\pi^+)^2}
+\frac{8}{3}\frac{\partial\theta^I_{\dot{c}}\partial\theta^{\dot{c}}_K
\partial\theta^K_{\dot{a}}}{(\pi^+)^2}, \label{6dDadot}\\
%%%%%%%%%%%%%%%%
\widehat{\mathcal{T}}=&\mathcal{T}
-\frac{1}{4}\frac{\partial^2\log\pi^+}{\pi^+}
-2\frac{\partial^2\theta^K_{\dot{c}}\partial\theta^{\dot{c}}_K}
{(\pi^+)^2}
-\frac{8}{3}\frac{\partial\theta^K_{\dot{c}}\partial\theta^{\dot{d}}_K
\partial\theta^L_{\dot{d}}\partial\theta^{\dot{c}}_L}{(\pi^+)^3},\label{6dT}
\end{align}
\end{subequations}
and satisfy
\begin{subequations}
\begin{align}
\widehat{D}^I_{\dot{a}}(z)\widehat{D}^J_{\dot{b}}(w)\sim&
-\frac{4\epsilon^{IJ}\epsilon_{\dot{a}\dot{b}}\widehat{\mathcal{T}}(w)}{z-w},\\
\textrm{rest}\sim&0.
\end{align}
\end{subequations}

The BRST charge is straitforwardly constructed from this
constraint algebra as
\begin{equation}
\tilde{Q}=\oint\dz\left(
\tilde{\lambda}^\alpha_I\widehat{D}^I_\alpha
+c\widehat{\mathcal{T}}
-2\tilde{\lambda}^I_{\dot{a}}\tilde{\lambda}^{\dot{a}}_Ib\right),
\label{BRSilde6}
\end{equation}
with the unconstrained bosonic ghost $\tilde{\lambda}^\alpha_I$
and the fermionic ghost pair $b$ and $c$ with $b(z)c(w)\sim 1/(z-w)$.
This BRST charge (\ref{BRSilde6}) is exactly nilpotent.

We can define the composite $B$-field as
\begin{equation}
 B=b\pi^+-\tilde{\omega}^I_\alpha\partial\theta^\alpha_I,
\label{b6}
\end{equation}
where $\tilde{\omega}^I_\alpha$ is the conjugate bosonic
anti-ghost satisfying
\begin{equation}
 \tilde{\lambda}^\alpha_I(z)\tilde{\omega}^J_\beta(w)
\sim\frac{\delta_I^J\delta^\alpha_\beta}{z-w}.
\end{equation}
Then, the energy-momentum tensor can be obtained as
\begin{align}
\{\tilde{Q},B(z)\}=&T(z),\nonumber\\
T=&-\frac{1}{2}\pi^\mu\pi_\mu-\frac{1}{4}\partial^2\log\pi^+
-\frac{1}{2}S^I_a\partial S^a_I
-d^I_\alpha\partial\theta^\alpha_I
-\omega^I_\alpha\partial\lambda^\alpha_I
-b\partial c.
\end{align}
This energy-momentum tensor has vanishing central charge,
and the similarity transformation defined by
\begin{equation}
 R=-\oint\dz bc\log\pi^+
\end{equation}
yields the conventional one with the reparametrization ghosts
\begin{align}
e^RTe^{-R}=&
-\frac{1}{2}\pi^\mu\pi_\mu+\frac{3}{4}\partial^2\log\pi^+
-\frac{1}{2}S^I_a\partial S^a_I-d^I_\alpha\partial\theta^\alpha_I
-\omega^I_\alpha\partial\lambda^\alpha_I
-\partial bc-2b\partial c.
\end{align}

\subsection{Equivalence to the PS formalism in
six dimensions}

Before showing coincidence of the physical spectra of
the DS and PS formalisms,
we have to explicitly solve the pure-spinor constraint 
in six dimensions.
The pure-spinor constraint in six dimensions can be written
\begin{equation}
\lambda_1^\alpha(C\gamma^\mu)_{\alpha\beta}\lambda_2^\beta=0,
\label{psin6}
\end{equation}
where $\lambda_{1,2}^\alpha=\lambda_{I=1,2}^\alpha$.
This equation can be satisfied if $\lambda_2^\alpha$ is
proportional to $\lambda_1^\alpha$.\footnote{
As in the four-dimensional case, we consider
$\lambda^\alpha_1$ and $\lambda^\alpha_2$ to be
two independent Weyl spinors, which are actually
complex conjugates, due to the $SU(2)$-Majorana constraint (\ref{su2m}).}
Because the proportionality constant is arbitrary, the number of
independent degrees of freedom of a pure spinor $\lambda_I^\alpha$
becomes $8-3=5$. Equation (\ref{psin6}) can be explicitly solved 
by using the light-cone decomposition as follows. 
Let us first consider a component of (\ref{psin6}),
\begin{equation}
\lambda_1C\gamma^+\lambda_2=\lambda_{1\dot{a}}\lambda_2^{\dot{a}}=0.
\label{plus}
\end{equation}
If we introduce the spinor dual to $\lambda_1^{\dot{a}}$ as
$\lambda_{1\dot{a}}l^{\dot{a}}=1$, the arbitrary spinor can be expanded
in $\lambda_1^{\dot{a}}$ and $l^{\dot{a}}$ as
\begin{equation}
 \lambda_2^{\dot{a}}=\alpha\lambda_1^{\dot{a}}
+\tilde{\alpha} l^{\dot{a}},
\end{equation}
where the expansion coefficients are
\begin{equation}
 \alpha=\lambda_{2\dot{a}}l^{\dot{a}},\qquad
\tilde{\alpha}=-\lambda_{2\dot{a}}\lambda_1^{\dot{a}}.
\end{equation}
Because the constraint (\ref{plus}) leads to $\tilde{\alpha}=0$,
we have
\begin{equation}
\lambda_2^{\dot{a}}=\alpha\lambda_1^{\dot{a}}
\end{equation}
for the pure spinor. The next constraint,
\begin{align}
\lambda_1C\gamma^i\lambda_2=&
\lambda_1^{\dot{a}}(\epsilon\tilde\gamma^i)_{\dot{a}b}\lambda_2^b
-\lambda_2^{\dot{b}}(\epsilon\tilde\gamma^i)_{\dot{b}a}\lambda_1^a
\nonumber\\
=&\lambda_1^{\dot{a}}(\epsilon\tilde\gamma^i)_{\dot{a}b}
(\lambda_2^b-\alpha\lambda_1^b)=0,
\end{align}
is satisfied by
\begin{equation}
\lambda_2^a=\alpha\lambda_1^a,
\end{equation}
with which the final constraint, $\lambda_1C\gamma^-\lambda_2=0$,
is automatically satisfied. 
In summary, we can satisfy the pure-spinor constraint with
$\lambda_2^\alpha=\alpha\lambda_1^\alpha$, where
$\alpha=\lambda_{2\dot{a}}l^{\dot{a}}$ with
$\lambda_{1\dot{a}}l^{\dot{a}}=1$.

Now we can show that the cohomology of the BRST charge $\tilde{Q}$
(\ref{BRSilde6}) of the DS formalism coincides with that of 
the PS formalism (\ref{psbrs6}).
Using the first similarity transformation generated by
\begin{equation}
X=\frac{1}{4}\oint\frac{dz}{2\pi i}c(l^{\dot{a}}\widehat{D}^2_{\dot{a}}), 
\end{equation}
the BRST charge becomes
\begin{align}
&e^X\tilde{Q}e^{-X}=
\delta_b+Q^{(1)},\\
&\delta_b=-\oint\frac{dz}{2\pi i}4\tilde{\alpha}b,\\
&Q^{(1)}=\oint\frac{dz}{2\pi i}
(\tilde{\lambda}_I^a\widehat{D}^a_I+\tilde{\lambda}_1^{\dot{a}}
(\widehat{D}^1_{\dot{a}}+\alpha \widehat{D}^2_{\dot{a}})),
\end{align}
where $\alpha=\tilde{\lambda}_{2\dot{a}}l^{\dot{a}}$ and 
$\tilde{\alpha}=-\tilde{\lambda}_{2\dot{a}}\tilde{\lambda}_1^{\dot{a}}$.
Then $\tilde{\alpha},\tilde{\beta},c$ and $b$ are decoupled as a quartet,
where $\tilde{\beta}$ is the conjugate of $\tilde{\alpha}$ 
which can, in principle, be constructed from $\tilde{\omega}$.

The second similarity transformations, generated by
\begin{subequations}
\begin{align}
Y=&-\frac{1}{2}\oint\frac{dz}{2\pi i}S^1_aS^{2a}\log\pi^+,\\
Z^1=&\oint\frac{dz}{2\pi i}\left(\frac{1}{\sqrt{2}}\left(
S^{1a}+\frac{\alpha}{\pi^+}S^{2a}\right)d^2_a
+\frac{2}{\alpha}\frac{\partial\theta^{1\dot{a}}}{\pi^+}
(\partial\theta^1_{\dot{a}}+\alpha\partial\theta^2_{\dot{a}})\right),\\
Z^2=&\frac{\alpha}{2}\oint\frac{dz}{2\pi i}
\frac{S^2_aS^{2a}}{\pi^+},
\end{align}
\end{subequations}
give
\begin{align}
&e^{Z^2}e^{Z^1}e^YQ^{(1)}e^{-Y}e^{-Z^1}e^{-Z^2}=\delta+Q,\\
&\delta=\oint\frac{dz}{2\pi i}\sqrt{2}
(\tilde{\lambda}_2^a-\alpha\tilde{\lambda}_1^a)S^2_a,\\
&Q=\oint\frac{dz}{2\pi i}\left(
\tilde{\lambda}_1^a(d^1_a+\alpha d^2_a)
+\tilde{\lambda}_1^{\dot{a}}(d^1_{\dot{a}}+\alpha d^2_{\dot{a}})\right).
\label{Qfinal}
\end{align}
Then $(\tilde{\lambda}_2^a-\alpha\tilde{\lambda}_1^a),\
(\tilde{\omega}^2_a-\alpha\tilde{\omega}^1_a),\ S^1_a$ and $S^2_a$ 
are decoupled as a quartet.
The final form of $Q$ (\ref{Qfinal}) can be written as
\begin{equation}
Q=\oint\dz \tilde{\lambda}^\alpha_Id^I_\alpha, 
\end{equation}
with the constraint 
$\tilde{\lambda}_{2\alpha}=\alpha\tilde{\lambda}_{1\alpha}$,
which is the BRST charge of the PS formalism (\ref{psbrs6}).

\section{Coupling to the Calabi-Yau sector}\label{CY}

In the conventional formulations,
a lower-dimensional superstring is not consistent by
itself but must be combined with additional degrees
of freedom which come from the compactified space.
Here, since we assume the lower-dimensional supersymmetry,
it is known that such degrees of freedom are represented
by some unitary representations of $N=2$ superconformal
field theory, which we describe by using the generators 
$(T_C,G^\pm_C,J_C)$ satisfying
\begin{subequations}
\begin{align}
T_C(z)T_C(w)\sim&
\frac{c/2}{(z-w)^4}+\frac{2T_C(w)}{(z-w)^2}
+\frac{\partial T_C(w)}{z-w},\\
T_C(z)G^\pm_C(w)\sim&
\frac{\frac{3}{2}G^\pm_C(w)}{(z-w)^2}+\frac{\partial G^\pm_C(w)}{z-w},\qquad
T_C(z)J_C(w)\sim
\frac{J_C(w)}{(z-w)^2}+\frac{\partial J_C(w)}{z-w},\\
G^+_C(z)G^-_C(w)\sim&
\frac{c/3}{(z-w)^3}+\frac{J_C(w)}{(z-w)^2}
+\frac{T_C(w)+\frac{1}{2}\partial J_C(w)}{z-w}, \\
J_C(z)G^\pm_C(w)\sim&
\pm\frac{G^\pm_C(w)}{z-w},\qquad
J_C(z)J_C(w)\sim
\frac{c/3}{(z-w)^2},
\end{align}
\end{subequations}
where $c=9$ for $d=4$ and $c=6$ for $d=6$. 
We call this the Calabi-Yau (CY) sector.

In order to combine the lower-dimensional superstring
with this CY sector, the hidden topological 
$N=2$ superconformal symmetry\cite{BV,Ohta,OC} generated by
\begin{subequations}
\begin{align}
G^+=&J_{\textrm{BRST}}=\tilde{\lambda}^\alpha\widehat{D}_\alpha
+\tilde{\bar{\lambda}}^{\dot{\alpha}}
\widehat{\bar{D}}_{\dot{\alpha}}
+c\widehat{\mathcal{T}}
-4\tilde{\lambda^2}\tilde{\bar{\lambda}}^{\dot2}b,\\
%%%%%%%%%%%%%%%%%%%
G^-=&B=b\pi^+-\tilde{\omega}_\alpha\partial\theta^\alpha
-\tilde{\bar{\omega}}_{\dot{\alpha}}
\partial\bar{\theta}^{\dot{\alpha}},\\
%%%%%%%%%%%%%%%%%%%
T=&-\frac{1}{2}\pi^\mu\pi_\mu-\frac{1}{8}\partial^2\log\pi^+
-\frac{1}{2}S\partial\bar{S}+\frac{1}{2}\partial S\bar{S}
\nonumber\\
&-d_\alpha\partial\theta^\alpha
-\bar{d}_{\dot{\alpha}}\partial\bar{\theta}^{\dot{\alpha}}
-\tilde{\omega}_\alpha\partial\tilde{\lambda}^\alpha
-\tilde{\bar{\omega}}_{\dot{\alpha}}
\partial\tilde{\bar{\lambda}}^{\dot{\alpha}}-b\partial c,\\
%%%%%%%%%%%%%%%%%%%
J=&bc-\tilde{\lambda}^\alpha\tilde{\omega}_\alpha
-\tilde{\bar{\lambda}}^{\dot{\alpha}}
\tilde{\bar{\omega}}_{\dot{\alpha}},
\end{align}
\end{subequations}
for $d=4$, and
\begin{subequations}
\begin{align}
G^+=&J_{\textrm{BRST}}=\tilde{\lambda}^\alpha_I\widehat{D}_\alpha^I
+c\widehat{\mathcal{T}}
-2\tilde{\lambda}^I_{\dot{a}}\tilde{\lambda}^{\dot{a}}_Ib,\\
%%%%%%%%%%%%%%%%%%%
G^-=&B=b\pi^+-\tilde{\omega}^I_\alpha\partial\theta^\alpha_I,\\
%%%%%%%%%%%%%%%%%%%
T=&-\frac{1}{2}\pi^\mu\pi_\mu-\frac{1}{4}\partial^2\log\pi^+
-\frac{1}{2}S^I_a\partial S_I^a
-d_\alpha^I\partial\theta^\alpha_I
-\tilde{\omega}_\alpha^I\partial\tilde{\lambda}^\alpha_I
-b\partial c,\\
%%%%%%%%%%%%%%%%%%%
J=&bc-\tilde{\lambda}^\alpha_I\tilde{\omega}_\alpha^I,
\end{align}
\end{subequations}
for $d=6$, play an important role.
We can couple this topological superconformal algebra
(SCA) with the $N=2$ SCA in the CY sector by twisting it 
so as to obtain the topological SCA. 
This is consistent with the fact that the central charge of 
the lower-dimensional superstring is already zero, as explained
above. The energy-momentum tensor in the CY sector 
must also have vanishing central charge.
The BRST charge of the coupled system is then given by
\begin{equation}
 Q_B=\oint\dz\left(G^++G^+_C\right).\label{coupled}
\end{equation}
After applying the same similarity transformations as in the case 
without the CY sector, the cohomology is equivalent to that of the PS
formalism coupled to the topological string in the Calabi-Yau space 
with the BRST charge
\begin{equation}
 Q=Q_{PS}+\oint\dz G^+_C,
\label{topological}
\end{equation}
where $Q_{PS}$ is given by (\ref{psbrs4}) or (\ref{psbrs6}).
The four-dimensional case is treated in \citen{Ber}.

Another way to couple the lower-dimensional superstring to the CY sector
is by extending our constraint generators to those forming
the same algebra but including the generators of 
$N=2$ SCA in the CY sector. This is possible in the $d=4$ case if we choose
\begin{subequations}\label{cg4new}
\begin{align}
\widehat{D}_1=&D_1,\qquad
\widehat{\bar{D}}_{\dot1}=\bar{D}_{\dot1},\\
%%%%%%%%%%%%%%%%%%%%%%%%%%%%%%%
\widehat{D}_2=&D_2
+\frac{2}{\sqrt{\pi^+}}G^+_C+2\frac{\partial\bar{\theta}^{\dot2}J_C}{\pi^+}
-4\frac{\partial^2\bar{\theta}^{\dot2}}{\pi^+}
+2\frac{\partial\pi^+\partial\bar{\theta}^{\dot2}}{(\pi^+)^2},\\
%%%%%%%%%%%%%%%%%%%%%%%%%%%%%%%
\widehat{\bar{D}}_{\dot2}=&\bar{D}_{\dot2}
+\frac{2}{\sqrt{\pi^+}}G^-_C-2\frac{\partial\theta^2J_C}{\pi^+}
-4\frac{\partial^2\theta^2}{\pi^+}
+2\frac{\partial\pi^+\partial\theta^2}{(\pi^+)^2},\\
%%%%%%%%%%%%%%%%%%%%%%%%%%%%%%%
\widehat{\mathcal{T}}=&\mathcal{T}+\frac{T_C}{\pi^+}
+4\frac{\partial\theta^2\partial^2\bar{\theta}^{\dot2}}{(\pi^+)^2}
-4\frac{\partial^2\theta^2\partial\bar{\theta}^{\dot2}}{(\pi^+)^2}
-\frac{1}{2}\frac{\partial^2\log\pi^+}{\pi^+}.
\end{align}
\end{subequations}
The energy-momentum tensor then becomes
\begin{align}
 \{\tilde{Q},B(z)\}=&T(z),\nonumber\\
T=&-\frac{1}{2}\pi^\mu\pi_\mu-\frac{1}{2}\partial^2\log\pi^+
-\frac{1}{2}S\partial\bar{S}+\frac{1}{2}\partial S\bar{S}
\nonumber\\
&
-d_\alpha\partial\theta^\alpha
-\bar{d}_{\dot\alpha}\partial\bar{\theta}^{\dot\alpha}
-\omega_\alpha\partial\lambda^\alpha
-\bar{\omega}_{\dot\alpha}\partial\bar{\lambda}^{\dot\alpha}
+T_C-b\partial c.
\end{align}
It should be noted that the coefficient of the $\log\pi^+$
term is modified such that the total central charge vanishes.
From this form of the energy-momentum tensor, 
the lower-dimensional superstring seems to be coupled with
the CY sector without twisting.
However, if we modify the similarity
transformation as $Y\rightarrow Y+Y_C$ with
\begin{subequations} 
\begin{equation}
 Y_C=-\oint\dz J_C\log
\left(\frac{2\tilde{\lambda}^2}{\sqrt{\pi^+}}\right)
\end{equation}
or $\bar{Y}\rightarrow \bar{Y}+\bar{Y}_C$ with
\begin{equation}
 \bar{Y}_C=\oint\dz J_C\log
\left(\frac{2\tilde{\bar{\lambda}}^{\dot2}}{\sqrt{\pi^+}}\right),
\end{equation}
\end{subequations}
the total BRST charge becomes
\begin{subequations} 
\begin{equation}
Q^{tot}=Q+\oint\dz G^+, \label{cbrs1}
\end{equation} 
with the BRST charge $Q$ (\ref{brs4f1}), for the branch
$\bar{\lambda}^{\dot\alpha}=0$ and
\begin{equation}
Q^{tot}=Q+\oint\dz G^-, \label{cbrs2}
\end{equation} 
\end{subequations}
with the BRST charge $Q$ (\ref{brs4f2}), for the branch $\lambda^\alpha=0$.
The energy-momentum tensor in the CY sector is again twisted as
$T_C+\frac{1}{2}\partial J_C$ ($T_C-\frac{1}{2}\partial J_C$)
for the branch $\bar{\lambda}^{\dot\alpha}=0$ ($\lambda^\alpha=0$).
The cohomology of these BRST charges (\ref{cbrs1}) and (\ref{cbrs2}) 
are also essentially equivalent to that of the topological 
charge (\ref{topological}).\footnote{
The constraint algebra of the six-dimensional case can also be
extended if we impose the $N=4$ superconformal symmetry on
the CY sector. However, it seems there is no similarity transformation to relate it
to the PS formalism coupled with the CY sector.
}

\section{Discussion}\label{discuss}

In this paper we have investigated the lower-dimensional,
$d=4$ and $d=6$, superstrings using the DS formalism 
and shown that they are equivalent to the lower-dimensional PS
superstrings. The unexpected off-shell nature of
the physical spectrum can be interpreted as
a manifestation of  noncriticality in the PS formalism.

Because the lower-dimensional PS superstring 
has no Lorentz anomaly, the symmetry is also realized 
in the DS formalism, at least in the physical Hilbert space.
However, a direct study of 
Lorentz anomalies in the DS formalism is still worth carrying out,
as we have taken the semi-light cone gauge to quantize it. 
In fact, we have found that the physical Lorentz generator $M^{i-}$, 
which is non-trivial in the semi-light cone gauge,\cite{BM,GG} 
does not commute with the BRST charge in lower dimensions.
This result will be  reported in a forthcoming paper.\cite{KM}

It seems  peculiar that the off-shell
vector multiplet is included in the physical spectrum,
since the DS superstring is classically equivalent
with the GS superstring. However, this does not result in
a fatal contradiction, since the lower-dimensional
GS superstring is not well-defined,
due to the global Lorentz anomaly. In addition,
the massless spectrum of the GS superstring
in the light-cone gauge involves only half of 
the vector multiplet, which has only positive or 
negative helicities; there is yet no principle
that allows us to use two multiplets 
with both positive and negative
helicities simultaneously.
In any case, it is still mysterious why the off-shell
states appear as physical states, since in the initial stage
we have the Virasoro constraint, which comes from the world-sheet
reparametrization invariance. This should be clarified
in a future investigation.

It would be interesting to consider how to couple the 
lower-dimensional DS superstring with the compactified 
space degrees of freedom.
We have proposed two ways to combine the two sectors, although 
they are not completely successful.
It would also be interesting to study the relation 
to the hybrid formalism,\cite{hybrid}
and this may provide some information about how
to make the lower-dimensional superstring consistent.
It may also be possible for lower-dimensional superstrings 
to provide a framework to construct some \textit{off-shell} superstring 
which is different from the strings in the conventional string
theory.\cite{GKP,Mizo,AGMOY} It would be interesting 
to explore such a possibility.

\section*{Acknowledgements}

The authors would like to thank 
Yoichi Kazama for useful discussions.
The work of HK is supported in part by a Grant-in-Aid
for Scientific Research (No. 13135213) and a Grant-in-Aid
for the 21st Century COE \lq\lq Center for Diversity and
Universality in Physics,'' while  the work of SM is supported by 
a Grant-in-Aid for Scientific Research (No. 16540273)
from the Ministry of Education,
Culture, Sports, Science and Technology (MEXT) of Japan.

\appendix
\section{Notation and Conventions}\label{conventions}
%Empty argument \section{} yields `Appendix'. 

In this appendix we present our notation and conventions.
We use the metric $\eta^{\mu\nu}=\textrm{diag}(+1,-1,\cdots,-1)$.
The light-cone coordinate $x^\mu=(x^\pm,x^i)$ 
$(i=1,\cdots,d-2)$ is defined by
$x^\pm=x^0\pm x^{d-1}$. The gamma matrices satisfy
\begin{equation}
\left\{\Gamma^\mu,\Gamma^\nu\right\}=2\eta^{\mu\nu}.
\end{equation}
We use the two-(four-)component spinor notation for $d=4$ ($d=6$),
employing the following explicit representations of the gamma matrices. 

\subsection{$d=4$}

We use the two-component notation for the gamma matrices
\begin{equation}
\Gamma_\mu=
\begin{pmatrix}
 0&(\sigma_\mu)_{\alpha\dot{\beta}} \\
 (\bar\sigma_\mu)^{\dot{\alpha}\beta}&0 \\ 
\end{pmatrix},\qquad
\Gamma_5=i\gamma^0\gamma^1\gamma^2\gamma^3=
\begin{pmatrix}
 1  & 0 \\
 0 & -1 \\
\end{pmatrix},
\end{equation}
where $\alpha,\beta,\dot{\alpha},\dot{\beta}=1,2$ and
\begin{equation}
(\sigma_\mu)_{\alpha\dot{\beta}}=(1,\boldsymbol{\sigma}),\quad
(\bar\sigma_\mu)^{\dot{\alpha}\beta}=(1,-\boldsymbol{\sigma}).
\end{equation}
Here, $\boldsymbol{\sigma}=(\sigma_1,\sigma_2,\sigma_3)$ 
represents the Pauli matrices. 
These are related as
$(\bar{\sigma}^\mu)^{\dot{\alpha}\alpha}=
\epsilon^{\alpha\beta}\epsilon^{\dot{\alpha}\dot{\beta}}
(\sigma^\mu)_{\beta\dot{\beta}}$, 
where the two-dimensional antisymmetric 
two-spinor is defined by
$\epsilon^{\alpha\beta}= \epsilon_{\alpha\beta}
=\epsilon^{\dot{\alpha}\dot{\beta}}
=\epsilon_{\dot{\alpha}\dot{\beta}}=i\sigma_2$.
The useful Fierz identities are given by
\begin{alignat}{4}
(\sigma^\mu)_{\alpha\dot{\beta}}(\bar\sigma_\mu)^{\dot{\gamma}\delta}=&
2\delta_\alpha^\delta\delta_{\dot{\beta}}^{\dot{\gamma}},\qquad&
(\sigma^\mu)_{\alpha\dot{\beta}}(\sigma_\mu)_{\gamma\dot{\delta}}=&
2\epsilon_{\alpha\gamma}\epsilon_{\dot{\beta}\dot{\delta}}.
\end{alignat}

The indices of the two-component spinors are raised and lowered as
\begin{subequations}\label{rule}
\begin{alignat}{3}
\theta^\alpha=&\epsilon^{\alpha\beta}\theta_\beta,\qquad
\theta_\alpha=&\theta^\beta\epsilon_{\beta\alpha},\\
\bar{\theta}^{\dot{\alpha}}=&
\epsilon^{\dot{\alpha}\dot{\beta}}\bar{\theta}_{\dot{\beta}}, \qquad
\bar{\theta}_{\dot{\alpha}}=&
\bar{\theta}^{\dot{\beta}}\epsilon_{\dot{\beta}\dot{\alpha}}.
\end{alignat}
\end{subequations}
In this notation, a Majorana spinor is represented by the pair
of complex conjugate Weyl spinors $(\theta^\alpha)^*=\bar{\theta}^{\dot\alpha}$.

\subsection{$d=6$}

The four-component notation in six dimensions is that employing
the representation of $8\times 8$ gamma matrices
\begin{equation}
\Gamma_\mu=
\begin{pmatrix}
 0&\bar{\gamma}_\mu \\
 \gamma_\mu &0 \\ 
\end{pmatrix},\qquad
\Gamma_7=\Gamma_0\Gamma_1\Gamma_2\Gamma_3\Gamma_4\Gamma_5=
\begin{pmatrix}
 1 & 0\\
0 & -1\\
\end{pmatrix},\label{gamma}
\end{equation}
where 
\begin{equation}
\gamma_\mu=(\gamma_0,\gamma_i,\gamma_5),\quad
\bar{\gamma}_\mu=(\gamma_0,-\gamma_i,-\gamma_5)\qquad
(i=1,\cdots,4)
\end{equation} are $4\times4$ matrices whose forms we choose
as follows:
\begin{alignat}{3}\label{grep6}
 \gamma_0=&1\otimes 1=
\begin{pmatrix}
1 & 0 \\
0 & 1 \\ 
\end{pmatrix},&
\gamma_1=&\sigma_2\otimes\sigma_1=
\begin{pmatrix}
 0 & -i\sigma_1 \\
 i\sigma_1 & 0 \\
\end{pmatrix},
\nonumber\\
%%%%%%%%%%%%%%%%%%
\gamma_2=&\sigma_2\otimes\sigma_2=
\begin{pmatrix}
 0 & -i\sigma_2 \\
 i\sigma_2 & 0 \\
\end{pmatrix},\qquad &
\gamma_3=&\sigma_2\otimes\sigma_3=
\begin{pmatrix}
 0 & -i\sigma_3 \\
 i\sigma_3 & 0 \\
\end{pmatrix},
\nonumber\\
%%%%%%%%%%%%%%%%%%%
\gamma_4=&\sigma_1\otimes 1_2=
\begin{pmatrix}
0 & 1 \\
1 & 0\\ 
\end{pmatrix},&
\gamma_5=&\sigma_3\otimes 1_2=
\begin{pmatrix}
1 & 0 \\
0 & -1 \\ 
\end{pmatrix}.
\end{alignat}

The charge conjugation matrix satisfying
\begin{align}
\Gamma^{\mu T}=&-\mathcal{C}\Gamma^\mu \mathcal{C}^{-1},
\label{bigC}\\
\mathcal{C}^T=&\mathcal{C}
\end{align}
is given in this representation by
\begin{align}
\mathcal{C}=&\Gamma^2\Gamma^4\Gamma^5=i\sigma_2\otimes 1_2\otimes
 i\sigma_2,
\nonumber\\
=&
 \begin{pmatrix}
 0 & C \\
 C^T & 0
 \end{pmatrix},
\end{align}
with
\begin{align}
C=&1_2\otimes i\sigma_2,\nonumber\\
 =&
\begin{pmatrix}
i\sigma_2 & 0\\
0 & i\sigma_2 
\end{pmatrix}
=-C^T.
\end{align}
From (\ref{bigC}), 
$\gamma^\mu$ and $\bar{\gamma}^\mu$ satisfy
\begin{equation}
(\gamma^\mu)^T=C\gamma^\mu C^{-1},\qquad 
(\bar{\gamma}^\mu)^T=C\bar{\gamma}^\mu C^{-1}.
\end{equation}
Using this charge conjugation matrix,
it is natural to consider the combinations
\begin{align}
 \mathcal{C}\Gamma_\mu=&
\begin{pmatrix}
(C\gamma_\mu)_{\alpha\beta} & 0\\
0 & -(C\bar{\gamma_\mu})^{\dot{\alpha}\dot{\beta}} 
\end{pmatrix},\\
\Gamma_\mu\mathcal{C}^{-1}=&
\begin{pmatrix}
(\bar{\gamma}_\mu C^{-1})^{\alpha\beta} & 0\\
0 & -(\gamma_\mu C^{-1})_{\dot{\alpha}\dot{\beta}} 
\end{pmatrix}.
\end{align}
All of these $4\times4$ matrices are anti-symmetric,
\textit{i.e.} 
\begin{subequations}
\begin{alignat}{4}
(C\gamma^\mu)_{\alpha\beta}=&-(C\gamma^\mu)_{\beta\alpha}, \qquad&
(C\bar{\gamma}^\mu)^{\dot{\alpha}\dot{\beta}}
=&-(C\bar{\gamma}^\mu)^{\dot{\beta}\dot{\alpha}},\\
%%%%%%%%%%%%%%%%%%%%%%%%%%%%%%%%%%
(\bar{\gamma}^\mu C^{-1})^{\alpha\beta}
=&-(\bar{\gamma}^\mu C^{-1})^{\beta\alpha},\qquad &
(\gamma^\mu C^{-1})_{\dot{\alpha}\dot{\beta}}=&
-(\gamma^\mu C^{-1})_{\dot{\beta}\dot{\alpha}},
\end{alignat}
\end{subequations}
and are related as
\begin{equation}
\frac{1}{2}\epsilon^{\alpha\beta\gamma\delta}
(C\gamma^\mu)_{\gamma\delta}=-(\bar{\gamma}^\mu C^{-1})^{\alpha\beta},\qquad
\frac{1}{2}\epsilon_{\alpha\beta\gamma\delta}
(\bar{\gamma^\mu}C^{-1})^{\gamma\delta}=-(C\gamma^\mu)_{\alpha\beta}.
\label{dual}
\end{equation}

An important Fierz identity is given by
\begin{equation}
\delta^{[\alpha}_\gamma\delta^{\beta]}_\delta=
\frac{1}{2}(\delta^\alpha_\gamma\delta^\beta_\delta-
\delta^\beta_\gamma\delta^\alpha_\delta)=
-\frac{1}{4}(C\gamma^\mu)_{\gamma\delta}(\bar{\gamma}_\mu C^{-1})^{\alpha\beta},
\end{equation}
which can be rewritten using Eq.~(\ref{dual}) as
\begin{equation}
(C\gamma_\mu)_{\alpha\beta}(C\gamma^\mu)_{\gamma\delta}
=2\epsilon_{\alpha\beta\gamma\delta},\qquad
(\bar{\gamma}_\mu C^{-1})^{\alpha\beta}
(\bar{\gamma}^\mu C^{-1})^{\gamma\delta}=2\epsilon^{\alpha\beta\gamma\delta}, 
\end{equation}
or equivalently
\begin{equation}
(C\gamma_\mu)_{\alpha(\beta}(C\gamma^\mu)_{\gamma)\delta}=0,\qquad
(\bar{\gamma}_\mu C^{-1})^{\alpha(\beta} (\bar{\gamma}^\mu C^{-1})^{\gamma)\delta}=0. 
\end{equation}

The $SU(2)$-Majorana-Weyl (MW) spinor $\theta_{+I}\ (I=1,2)$ satisfies
the conditions
\begin{subequations}
\begin{align}
\Gamma_7\theta_{+I}=\theta_{+I},\\
\bar{\theta}_+^I=\epsilon^{IJ}\theta^T_{+J}\mathcal{C},\label{su2m}
\end{align}
\end{subequations}
where $\epsilon^{IJ}$ is the $SU(2)$ anti-symmetric tensor
defined by $\epsilon^{12}=\epsilon_{12}=1$, which is used to
raise or lower the index $I$ using the same rule as
in the case of the four-dimensional spinor (\ref{rule}).
This $SU(2)$ MW spinor can be written in the four-component 
notation as
\begin{equation}
 \theta_{+I}=
\begin{pmatrix}
\theta_I^\alpha\\
0\\ 
\end{pmatrix},\qquad
(\theta_I^\alpha)^\dag=\epsilon^{IJ}(\theta_J^TC)^{\dot{\alpha}}.
\end{equation}
We can further decompose it with the light-cone decomposition
\begin{equation}
\theta_I^\alpha=
\begin{pmatrix}
\theta^a_I\\
\theta^{\dot{a}}_I\\ 
\end{pmatrix},\qquad (a,\dot{a}=1,2)
\end{equation}
where $a$ and $\dot{a}$ are the spinor indices of 
the transverse rotation $SO(4)\sim SU(2)\times SU(2)$.
The $SU(2)$ MW condition is now simply given by
\begin{subequations}
\begin{align}
(\theta^a_I)^*=&\epsilon^{IJ}\theta^b_J\epsilon_{ba}\equiv \theta^I_a,\\
(\theta^{\dot{a}}_I)^*=&\epsilon^{IJ}\theta^{\dot{b}}_J\epsilon_{\dot{b}\dot{a}}
\equiv \theta^I_{\dot{a}}. 
\end{align}
\end{subequations}

From the representation (\ref{grep6}), the transverse components
of the gamma matrices can also be decomposed using the $2\times2$ matrices
$\tilde{\gamma}_i$ defined by
\begin{equation}
\tilde\gamma_i=i\sigma_i\quad (i=1,2,3),\qquad
\tilde\gamma_4=1_2.
\end{equation}
It is also useful to define $\tilde{\bar{\gamma}}_i$ as
\begin{equation}
\tilde{\bar{\gamma}}_i=-i\sigma_i\quad (i=1,2,3),\qquad
\tilde{\bar{\gamma}}_4=1_2.
\end{equation}
These two  $2\times2$ matrices are related as
\begin{equation}
 (\epsilon\tilde{\gamma}_i)_{\dot{a}b}=
-(\epsilon\tilde{\bar{\gamma}}_i)_{b\dot{a}},
\end{equation}
where $\epsilon=i\sigma_2$.

The two quantities $\bbsla{\Pi}\ $ and $\bbsla{\bar{\Pi}}\ $ 
constitute a representation of the quaternion and its conjugate,
and they satisfy the relation
\begin{equation}
 \bbsla{\Pi}\ \bbsla{\bar{\Pi}}=
 \bbsla{\bar{\Pi}}\ \bbsla{\Pi}=\Pi^i\Pi^i. 
\end{equation}

%\section{Second Appendix}

\end{document}